\documentclass[journal]{IEEEtran}
\usepackage{amsmath, amsfonts}
\usepackage{amsthm}
\usepackage{indentfirst}
\usepackage{caption}
\usepackage[utf8]{inputenc}
\usepackage{svg}
\usepackage{xcolor}
\usepackage[caption=false,font=footnotesize]{subfig} 
\usepackage{amssymb}
\usepackage{algorithmic}
\usepackage{algorithm}
\usepackage{array}
\usepackage{booktabs}
\usepackage{textcomp}
\usepackage{bm}
\usepackage{stfloats}
\usepackage{url}
\usepackage{verbatim}
\usepackage{graphicx}
\usepackage{hyperref}
\usepackage{cite}
\usepackage{afterpage}
\newtheorem{theorem}{Theorem}

\begin{document}

\title{Tracking-Assisted Robust Secure Transmission Against\\ a Mobile Eavesdropper in Cell-Free ISAC Networks}

\author{Jong-Hyuk Hong, Chaedam Son, and Si-Hyeon Lee,~\IEEEmembership{Senior Member,~IEEE}
\thanks{The authors are with the School of Electrical Engineering, Korea Advanced Institute of Science and Technology (KAIST), Daejeon 34141, South Korea (e-mail: jonghyukm41@kaist.ac.kr; scd5929@kaist.ac.kr; sihyeon@kaist.ac.kr). \emph{(Corresponding author: Si-Hyeon Lee.)}}}%

\setlength{\arraycolsep}{5pt}    
\renewcommand{\arraystretch}{0.9}

\maketitle
\begin{abstract}
In this paper, we propose a tracking-assisted robust secure transmission framework for cell-free integrated sensing and communication (ISAC) networks that exploits distributed multistatic sensing to recursively track a mobile eavesdropper and quantify the associated position uncertainty. Since robust beamforming requires communication channel uncertainty rather than position uncertainty, directly incorporating tracking information into secure transmission is nontrivial. This challenge becomes more pronounced in cell-free ISAC, where the common position uncertainty propagates differently to the channels of geographically distributed access points (APs), resulting in coupled AP-specific channel uncertainties. To address this challenge, we fuse multistatic sensing measurements from distributed APs and sensing receivers using an extended Kalman filter (EKF) and propose a Jacobian-based anisotropic ellipsoidal uncertainty model that maps the EKF position-error covariance to coupled AP-specific channel uncertainties. Based on the proposed model, we formulate a robust sum secrecy-rate maximization problem under per-AP transmit-power and sensing mean-square error constraints and develop an efficient alternating optimization algorithm for the joint design of communication beamforming and sensing signals. Simulation results demonstrate a fundamental trade-off between tracking accuracy and secrecy performance, the benefits of distributed multistatic sensing and cooperative transmission in cell-free ISAC, and the effectiveness of the proposed robust design in improving secrecy reliability under mobility-induced channel uncertainty.

\end{abstract}


\section{Introduction}

\IEEEPARstart{I}{ntegrated} sensing and communication (ISAC) has emerged as a key enabling technology for beyond fifth-generation and sixth-generation wireless networks, driven by the growing demand for high-quality communication and accurate environmental awareness~\cite{liu2022survey}. By jointly designing sensing and communication functionalities within a unified framework, ISAC enables efficient reuse of spectrum, hardware, infrastructure, and signal resources, thereby improving spectral efficiency, resource utilization, and system-level integration~\cite{liu2022integrated}. Furthermore, the sensing capability of ISAC can enhance communication by providing environmental awareness that enables adaptive transmission design~\cite{zhang2021enabling}.

Such environmental awareness is also valuable for communication security. Physical-layer security (PLS) provides an effective means of preventing information leakage to potential eavesdroppers by exploiting the intrinsic characteristics of wireless channels without relying solely on conventional cryptographic techniques~\cite{liu2026secure}. However, the locations and channels of potential eavesdroppers are generally unknown to the transmitter, making it challenging to accurately suppress information leakage. In this context, ISAC can naturally support PLS by sensing potential eavesdroppers and incorporating the acquired information into secure transmission design. However, because communication and sensing share the same transmit resources, secure transmission and eavesdropper sensing become tightly coupled, making their joint optimization a fundamental challenge in secure ISAC systems~\cite{wu2018survey}.

To address this challenge, extensive efforts have been devoted to sensing-assisted secure ISAC designs~\cite{son2024secrecy,ren2023robust,su2023sensing}.
In~\cite{son2024secrecy}, a secure ISAC system enabled by an autonomous aerial vehicle was proposed, in which beamforming and the vehicle trajectory were jointly optimized to maximize the sum secrecy rate. In~\cite{ren2023robust}, robust beamforming was developed by exploiting dedicated sensing signals as artificial noise against potential eavesdroppers. In~\cite{su2023sensing}, sensing-assisted secure transmission was investigated by incorporating the estimated eavesdropper directions into beamforming design.

However, the above studies mainly consider static eavesdroppers or assume that their instantaneous spatial information is available for secure transmission design. In practice, an eavesdropper may move continuously, causing its position and associated channels to vary over time. Consequently, instantaneous sensing alone is insufficient for reliable secure transmission, necessitating recursive state tracking that continuously predicts and updates the eavesdropper state while quantifying the associated estimation uncertainty.
Among various tracking techniques, the extended Kalman filter (EKF) has been widely adopted for mobile-eavesdropper tracking because it recursively estimates dynamic states from nonlinear sensing measurements while simultaneously quantifying the corresponding estimation uncertainty~\cite{liu2023securing,wei2023integrated,xu2025sensing}. Existing studies have exploited the EKF for trajectory prediction, channel state information (CSI) acquisition, and robust secure transmission against mobile eavesdroppers~\cite{liu2023securing,wei2023integrated,xu2025sensing}. Specifically,~\cite{liu2023securing} jointly predicted the trajectory and CSI of a mobile aerial eavesdropper for radar-assisted beamforming,~\cite{wei2023integrated} incorporated EKF-based state estimation into wiretap-channel prediction and robust resource allocation, and~\cite{xu2025sensing} tracked the angles, distances, and velocities of mobile eavesdroppers for secure transmission design.
Despite these advances, the tracking accuracy fundamentally depends on the quality and geometry of the available sensing measurements. As a mobile eavesdropper moves across the coverage area, favorable sensing links may disappear, resulting in weaker measurements and reduced geometric diversity, which ultimately degrades the tracking accuracy and the resulting secure transmission performance.  

To overcome this limitation, cell-free ISAC provides a promising architecture through geographically distributed access points (APs) and sensing receivers (SRs). The distributed network nodes collectively observe a mobile eavesdropper from multiple spatial perspectives, generating rich multistatic sensing measurements with enhanced geometric diversity and thereby enabling more reliable tracking across the coverage area. At the same time, cooperative transmission among the APs provides macro-diversity and abundant spatial degrees of freedom for secure communication~\cite{son2026robust}. These complementary sensing and communication capabilities make cell-free ISAC particularly attractive for secure transmission against mobile eavesdroppers. Despite these advantages, existing secure cell-free ISAC studies have primarily focused on static eavesdroppers or perfect eavesdropper information~\cite{kim2025fronthaul,nasir2024joint,ren2024secure}. Consequently, robust secure transmission that jointly exploits recursive mobile-eavesdropper tracking while accounting for the resulting time-varying channel uncertainty remains largely unexplored in cell-free ISAC systems.


Although EKF-based mobile-eavesdropper tracking and robust secure transmission have been extensively studied, integrating them in cell-free ISAC is fundamentally nontrivial. Unlike conventional robust beamforming, the EKF provides the uncertainty of the eavesdropper position rather than that of the communication channels. In a cell-free architecture, this common position uncertainty propagates differently to the line-of-sight (LoS) channels of geographically distributed APs, resulting in coupled AP-specific channel uncertainties. Consequently, the tracking uncertainty cannot be directly incorporated into secure transmission design, requiring a unified framework that bridges multistatic sensing, uncertainty modeling, and robust beamforming.


The main contributions of this paper are summarized as follows.
\begin{itemize}

\item We develop a mobility-aware secure cell-free ISAC framework for jointly tracking and suppressing a mobile eavesdropper. The proposed framework integrates multistatic sensing, EKF-based recursive tracking, and cooperative secure transmission through central processing unit (CPU)-level joint processing, thereby extending existing secure cell-free ISAC design from static to mobile eavesdropper scenarios.

\item We propose a geometry-aware uncertainty model for the AP-specific eavesdropper channels. Specifically, we develop a Jacobian-based anisotropic ellipsoidal uncertainty model that transforms the EKF position-error covariance into coupled AP-specific LoS channel uncertainties. The proposed model captures the geometric relationship between the common position uncertainty and distributed AP channels, enabling tractable robust beamforming.

\item Based on the proposed uncertainty model, we formulate a robust sum secrecy-rate maximization problem under per-AP transmit-power and EKF-based sensing mean-square error (MSE) constraints. We further develop an efficient alternating optimization (AO)-based algorithm that jointly optimizes communication beamforming and sensing signals while accounting for the worst-case eavesdropper channel uncertainty.

\item Simulation results demonstrate that the proposed framework consistently outperforms representative structural baselines in both secrecy performance and tracking accuracy. The results further reveal the benefits of distributed multistatic sensing and provide insights into the fundamental trade-offs among tracking accuracy, robustness, and secrecy performance.
\end{itemize}

{\it Notation}: Scalars, vectors, matrices, and sets are denoted by regular lowercase (or uppercase) letters, bold lowercase letters, bold uppercase letters, and calligraphic letters, respectively, i.e., $a$ (or $A$), ${\bf a}$, ${\bf A}$, and $\mathcal{A}$. 
The expectation operator is denoted by $\mathbb{E}[\cdot]$. 
$\mathbb{C}$ and $\mathbb{R}$ represent the sets of complex-valued and real-valued numbers, respectively. $j$ denotes the imaginary unit.
The operators $[\cdot]^T$, $[\cdot]^H$, $[\cdot]^*$, and $[\cdot]^{-1}$ denote transpose, Hermitian transpose, complex conjugate, and matrix inversion, respectively, and $\Re\{\cdot\}$ denotes the real part. 
For a scalar $a$, $|a|$ denotes the absolute value. 
The notation $\|\cdot\|$ denotes the Euclidean norm. 
$\mathcal{A}\setminus\{a\}$ denotes the set excluding $a$. 
${\bf I}_N$ and ${\bf 0}_N$ denote the $N\times N$ identity matrix and all-zero matrix, respectively. 
$\text{diag}(\cdot)$ and $\text{blkdiag}(\cdot)$ denote diagonal and block-diagonal matrices, respectively. 
$[{\bf A}]_{i,j}$ denotes the $(i,j)$-th element of matrix ${\bf A}$. 
The notation ${\bf A}\succeq {\bf 0}$ indicates that ${\bf A}$ is positive semidefinite. 
$\mathrm{Tr}(\cdot)$ denotes the trace operator.
\begin{figure}
    \centering
    \includegraphics[width=0.8\linewidth]{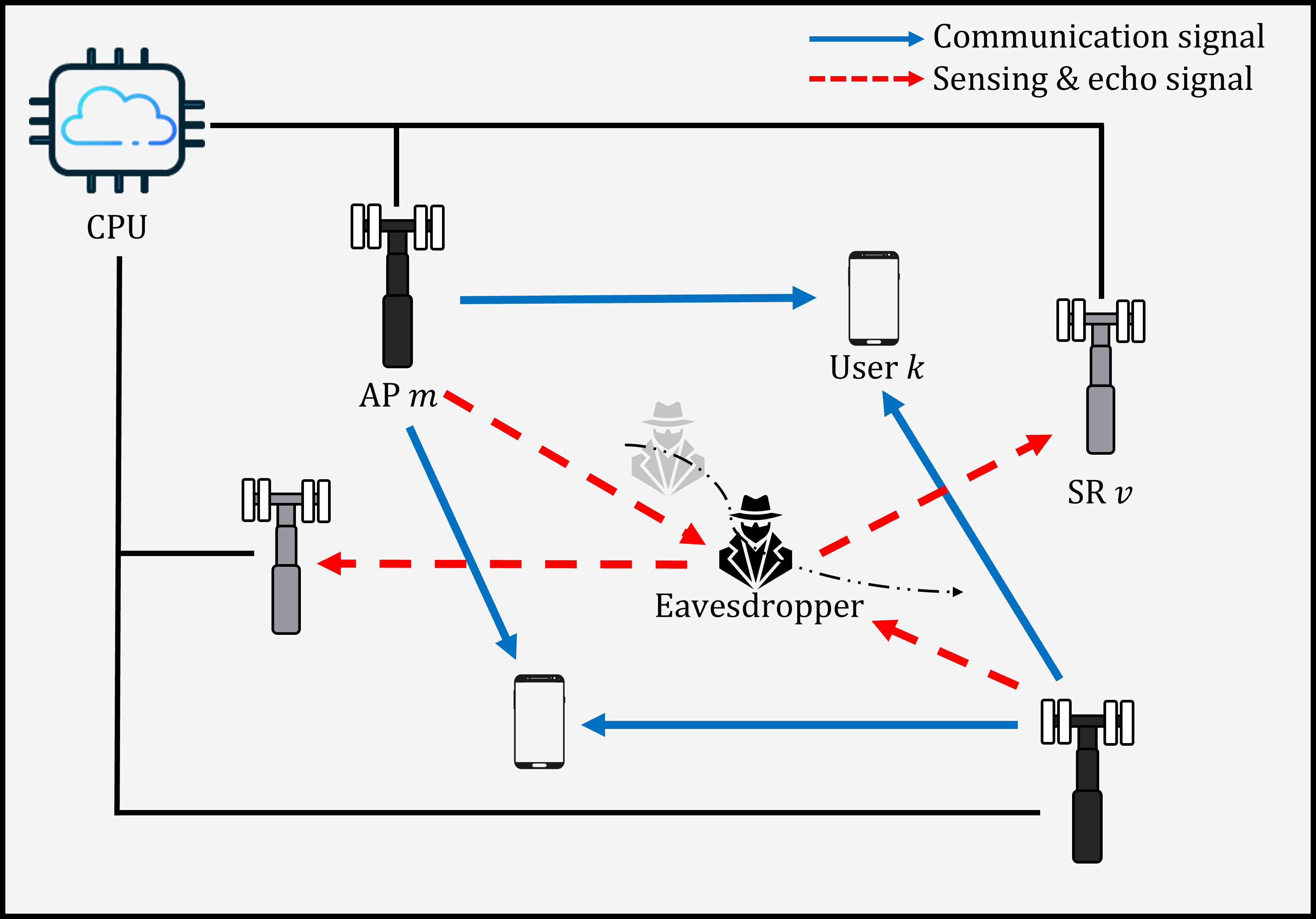}
    \caption{A cell-free ISAC network with a mobile eavesdropper.}
    \label{fig:sysmo}
\end{figure}
\section{System Model}
\label{sec:system_model}
As illustrated in Fig.~\ref{fig:sysmo}, we consider a secure cell-free ISAC system in which distributed APs provide downlink communication services to legitimate users while tracking a mobile eavesdropper.
Specifically, $M$ APs jointly serve $K$ single-antenna legitimate users, while $V$ SRs are deployed to receive the sensing echoes reflected from the eavesdropper. 
{All APs and SRs are connected to a CPU via ideal backhaul links with sufficient capacity~\cite{ngo2017cell} and are equipped with $N_t$-element and $N_r$-element uniform linear arrays (ULAs), respectively. The ULAs are assumed to have half-wavelength antenna spacing, i.e., $d=\lambda/2$, 
where $d$ and $\lambda$ denote the antenna spacing and carrier wavelength, respectively.}
Let $\mathcal{M}=\{1,2,\dots,M\}$, $\mathcal{V}=\{1,2,\dots,V\}$, and $\mathcal{K}=\{1,2,\dots,K\}$ denote the index sets of the APs, SRs, and legitimate users, respectively, and let $e$ denote the index of the eavesdropper. 

We consider an observation interval of duration $T$, which is divided into $L$ time slots indexed by $\mathcal{L}=\{1,2,\dots,L\}$, with slot duration $\Delta t$.
The legitimate users are assumed to remain static during the observation interval $T$.
In contrast, the eavesdropper is assumed to be mobile during the interval $T$, representing an adversarial scenario in which the eavesdropper changes its position to avoid persistent detection or tracking.
The slot duration $\Delta t$ is assumed to be sufficiently short so that the eavesdropper state, which consists of its position and velocity, can be regarded as constant over each time slot $l\in\mathcal{L}$ for communication and sensing signal modeling.

\subsection{Channel Model}
\label{sec:channel_model}
\subsubsection{Communication Channel}
The system is assumed to operate in a high-frequency band, where the LoS component typically dominates the propagation channel and the non-line-of-sight components are much weaker~\cite{nlos_mmwave1}. Therefore, only the LoS component is considered in this work. 
The channel between AP $m$ and $u\in\{e\}\cup\mathcal{K}$ at time slot $l$ is modeled as
\begin{equation}
    \mathbf{h}_{m,u}[l] =
    \sqrt{\frac{N_t\beta_0}{d_{m,u}^2[l]}}
    {\bf a}_t(\theta_{m,u}[l]),
    \label{eq:ch_exp}
\end{equation}
where $\beta_0$ denotes the channel power gain at a reference distance of $1\,\mathrm{m}$, $d_{m,u}[l]$ is the propagation distance between AP $m$ and node $u$, and $\theta_{m,u}[l]$ denotes the angle of departure (AoD) from AP $m$ toward node $u$. 
The transmit steering vector ${\bf a}_t(\theta)$ is given by
\begin{equation}
{\bf a}_t(\theta)
=
\frac{1}{\sqrt{N_t}}
\begin{bmatrix}
1,
e^{-j\frac{2\pi}{\lambda}d\sin\theta},
\dots,
e^{-j\frac{2\pi}{\lambda}d(N_t-1)\sin\theta}
\end{bmatrix}^T.
\end{equation}
Accordingly, we define the aggregate channel vector from all APs to node $u$ in the $l$-th time slot as
\begin{equation}
\mathbf{h}_u[l] \triangleq
\left[
\mathbf{h}_{1,u}^T[l],
\mathbf{h}_{2,u}^T[l],
\dots,
\mathbf{h}_{M,u}^T[l]
\right]^T
\in \mathbb{C}^{MN_t \times 1}.
\end{equation}
Since the legitimate users are assumed to be static and cooperative, their perfect CSI is assumed to be available throughout the transmission period. 
For notational simplicity, the time-slot index $l$ is omitted from ${\bf h}_k[l]$ for all $k\in\mathcal{K}$.
In contrast, since the eavesdropper is mobile and non-cooperative, its CSI is unavailable. Instead, the proposed sensing framework tracks the eavesdropper trajectory and predicts its {position}, based on which a nominal eavesdropper channel is constructed. The resulting position estimation error leads to channel uncertainty, whose modeling is presented in Section~\ref{sec:Uncertainty}.
\subsubsection{Sensing Channel}

For the AP--eavesdropper--SR bistatic sensing link, the eavesdropper is modeled as a point target~\cite{jscb}. Higher-order reflections are neglected due to the severe attenuation incurred by multiple-bounce propagation~\cite{sen_mult_neg}. Under these assumptions, the bistatic sensing channel from AP $m$ to SR $v$ via the eavesdropper at time slot $l$ is modeled as
\begin{equation}
{\bf H}_{m,v}[l]
=
\sqrt{
\frac{N_tN_r\beta_0 \rho_0 }
{ d_{m,e}^2[l] d_{v,e}^2[l] }
}
{\bf a}_r(\theta_{v,e}[l]){\bf a}_t^H(\theta_{m,e}[l]),
\end{equation}
where $\rho_0$ is the radar cross section of the eavesdropper, $d_{v,e}[l]$ is the distance between SR $v$ and the eavesdropper, and $\theta_{v,e}[l]$ is the corresponding angle of arrival (AoA) at SR $v$. The receive steering vector ${\bf a}_r(\theta)$ is given by
\begin{equation}
{\bf a}_r(\theta)
=
\frac{1}{\sqrt{N_r}}
\begin{bmatrix}
1,
e^{-j\frac{2\pi}{\lambda}d\sin\theta},
\dots,
e^{-j\frac{2\pi}{\lambda}d(N_r-1)\sin\theta}
\end{bmatrix}^T.
\end{equation}
\subsection{Signal Model}
\subsubsection{Transmit Signal}
\begin{figure*}
    \centering
    \includegraphics[width=0.7\linewidth]{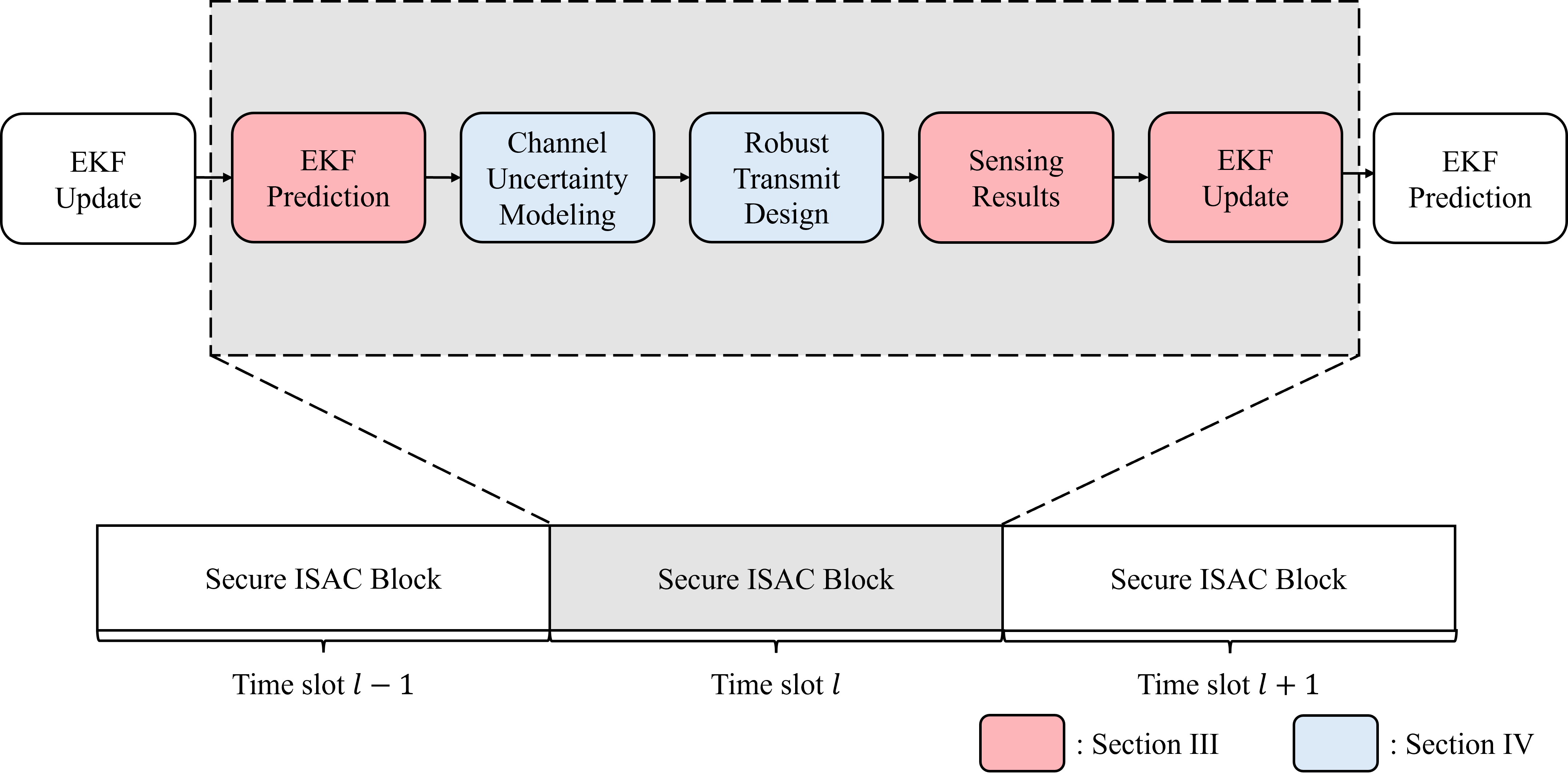}
    \caption{Proposed tracking-assisted robust secure transmission framework for cell-free ISAC.}
    \label{fig:protocol}
\end{figure*}
At time slot $l$, the transmit signal from AP $m$ is given by
\begin{equation}
\mathbf{x}_{m}[l]
= {\bf s}_{m}[l]+\sum_{k\in\mathcal{K}} \mathbf{w}_{m,k}[l] b_k[l],
\end{equation}
where $\mathbf{w}_{m,k}[l]\in\mathbb{C}^{N_t\times 1}$ denotes the beamforming vector from AP $m$ to user $k$, $\mathbf{s}_{m}[l]\in\mathbb{C}^{N_t\times 1}$ denotes the dedicated sensing signal transmitted by AP $m$, and $b_k[l]$ denotes the information symbol intended for user $k$. 
The symbol $b_k[l]$ is cooperatively transmitted by all APs to user $k$. The covariance matrix of $\mathbf{s}_{m}[l]$ is defined as ${\bf S}_{m}[l]=\mathbb{E}[{\bf s}_{m}[l]{\bf s}_{m}^H[l]]$, and the sensing signals transmitted by different APs are assumed to be mutually orthogonal, i.e., $\mathbb{E}[{\bf s}_{m}[l]{\bf s}_{m'}^H[l]]={\bf 0}_{N_t}$ for $m\neq m'$. The communication symbols $\{b_k[l]\}_{k\in\mathcal{K}}$ are modeled as mutually uncorrelated circularly symmetric complex Gaussian random variables with unit power, i.e., {$b_k[l]\sim\mathcal{CN}(0,1),~\forall k\in\mathcal {K}$} and $\mathbb{E}[b_k[l]b_{k'}^*[l]]=0$ for $k\neq k'$. For notational convenience, we stack the beamforming vectors across all APs as $\mathbf{w}_k[l]\triangleq \big[\mathbf{w}_{1,k}^T[l],\mathbf{w}_{2,k}^T[l],\dots,\mathbf{w}_{M,k}^T[l]\big]^T \in\mathbb{C}^{MN_t\times 1}$. 
By stacking the sensing signals transmitted by all APs, we define
${\bf s}[l]\triangleq[{\bf s}_{1}^{T}[l],\dots,{\bf s}_{M}^{T}[l]]^{T}\in\mathbb{C}^{MN_t\times 1}$.
Accordingly, we define the overall sensing covariance matrix as
$
{\bf S}[l]
=
\mathbb{E}\left[{\bf s}[l]{\bf s}^{H}[l]\right]
=
\mathrm{blkdiag}\big({\bf S}_{1}[l],\dots,{\bf S}_{M}[l]\big).
$
\subsubsection{Received Communication Signal Model}
In time slot $l$, the received signal at user $k$ is written as
\begin{align}
y_k[l]
&= \sum_{m \in \mathcal{M}} \mathbf{h}_{m,k}^H \mathbf{x}_{m}[l] + n_k[l]\notag \\
&= \underbrace{\mathbf{h}_{k}^H \mathbf{w}_{k}[l] b_k[l]}_{\text{Desired signal}}
+ \underbrace{\sum_{k' \in \mathcal{K}\setminus\{k\}}
\mathbf{h}_{k}^H \mathbf{w}_{k'}[l] b_{k'}[l]}_{\text{Inter-user interference}}\notag \\
&\quad
{+ \underbrace{
{\bf h}_{k}^H{\bf s}[l]}_{\text{Sensing interference}}}
+ n_k[l],
\end{align}
where $n_k[l]\sim\mathcal{CN}(0,\sigma_k^2)$ denotes the receiver noise at user $k$. The received signal at the eavesdropper in time slot $l$ is given by
\begin{equation}
\begin{aligned}
y_e[l]
&=\sum_{m \in \mathcal{M}} \mathbf{h}_{m,e}^H[l] \mathbf{x}_{m}[l] + n_e[l] \\
&= \sum_{k\in \mathcal{K}} \mathbf{h}_{e}^H[l] \mathbf{w}_{k}[l] b_{k}[l]
{+{\bf h}_{e}^H[l]{\bf s}[l]}
+ n_e[l],
\end{aligned}
\end{equation}
where $n_e[l]\sim\mathcal{CN}(0,\sigma_e^2)$ denotes the receiver noise at the eavesdropper. 
\subsubsection{Received Sensing Signal}
During the sensing process, the signals transmitted by the APs are received at the SRs through the direct AP--SR paths and reflections from the eavesdropper, static legitimate users, and environmental scatterers.
In the considered cell-free architecture, the transmitted waveforms and beamforming vectors are available to the network, while the direct AP--SR channels can be estimated, enabling direct-link interference cancellation during SR processing~\cite{jscb}.
Since the legitimate users are static and are not sensing targets, their reflections are assumed to be non-target static clutter and are suppressed together with reflections from static environmental scatterers~\cite{xiao2023integrated}.
Following the residual-interference model in~\cite{jscb}, the remaining components after direct-link interference cancellation and clutter suppression are modeled as zero-mean additive white Gaussian noise and incorporated into the effective noise.
Accordingly, the received sensing signal at SR $v$ is expressed as
\begin{equation}
\begin{aligned}
\mathbf{r}_{v}(l,t)\!
&=\!\!
\sum_{m\in\mathcal{M}}
e^{j2\pi\mu_{m,v}[l]t}
{\bf H}_{m,v}[l]
{\bf s}_{m}(l,t-\tau_{m,v}[l])
\!+\!
{\bf n}_{v}(l,t),
\end{aligned}
\label{eq:echo_pr}
\end{equation}
where $\tau_{m,v}[l]$ and $\mu_{m,v}[l]$ denote the propagation delay and Doppler frequency associated with the AP--eavesdropper--SR sensing link, respectively, and ${\bf n}_{v}(l,t)\sim\mathcal{CN}({\bf0},\sigma_v^2{\bf I}_{N_r})$ denotes the effective noise at SR $v$.

\subsection{Proposed Framework}
As described above, the legitimate-user channels remain static throughout the transmission period, whereas the eavesdropper state and channel evolve over time due to its mobility. To cope with this time-varying environment, we propose the EKF-based closed-loop framework illustrated in Fig.~\ref{fig:protocol}. At each time slot, the CPU first predicts the eavesdropper state and constructs the corresponding nominal channel. Based on the predicted state and its associated uncertainty, the transmit signal is then robustly designed for secure communication. After transmission, the SRs collect the sensing measurements and forward them to the CPU, which updates the eavesdropper state estimate via the EKF. The updated estimate is subsequently used to initialize the prediction in the next time slot, thereby closing the sensing--communication loop. The following sections present the main components of the proposed framework, including EKF-based eavesdropper trajectory tracking and uncertainty-aware robust transmit-signal design.
\section{Eavesdropper Trajectory Tracking}
\label{sec:tracking}
This section presents the EKF-based trajectory tracking method for estimating the time-varying state of the mobile eavesdropper. At each time slot, the EKF first predicts the eavesdropper state based on the previous estimate and then refines the prediction using the sensing measurements collected by the SRs. The resulting prediction and prediction covariance are subsequently used in Section~\ref{sec:Uncertainty} for nominal channel construction and position-induced channel uncertainty modeling.

Let ${\bf q}_{m}=[x_{m},y_{m}]^T$, ${\bf q}_{v}=[x_{v},y_{v}]^T$, and ${\bf q}_{e}[l]=[x_{e}[l],y_{e}[l]]^T$ denote the positions of AP $m$, SR $v$, and the eavesdropper at time slot $l$, respectively. The eavesdropper velocity is denoted by $\dot{\bf q}_{e}[l]=[\dot{x}_{e}[l],\dot{y}_{e}[l]]^T$. Accordingly, the EKF tracks the following eavesdropper state vector:
\[
{\bm \chi}[l]
=
\left[
{\bf q}_{e}^{T}[l],
\dot{\bf q}_{e}^{T}[l]
\right]^T.
\]

\subsection{State Evolution and EKF Prediction}
We adopt a nearly constant-velocity motion model for the eavesdropper state evolution, as commonly used in ISAC-based eavesdropper tracking~\cite{liu2023securing}:
\begin{equation}
{\bm\chi}[l] = {\bf F}{\bm \chi}[l-1]+{\bf n}_{\bm\chi}[l],
\quad
{\bf F}=
\begin{bmatrix}
{\bf I}_2 & \Delta t\cdot {\bf I}_2\\
{\bf 0}_2 & {\bf I}_2
\end{bmatrix},
\end{equation}
where ${\bf F}\in\mathbb{R}^{4\times 4}$ denotes the state transition matrix and
${\bf n}_{\bm\chi}[l]\sim\mathcal{N}\!\left({\bf 0},{\bf Q}_{{\bf n}_{\bm\chi}}\right)$ denotes the process noise. 
{We model the process-noise covariance as}
 $\mathbf{Q}_{{\bf n}_{\bm\chi}}
=
\mathrm{diag}\!\left(
\sigma_{x_e}^2,\,
\sigma_{y_e}^2,\,
\sigma_{\dot{x}_e}^2,\,
\sigma_{\dot{y}_e}^2
\right)$, which characterizes the uncertainty arising from unmodeled variations in the eavesdropper motion. 

{Given the state estimate $\hat{\bm\chi}[l-1]$ and its covariance matrix 
${\bf \Sigma}[l-1]$ obtained at time slot $l-1$, the EKF predicts the 
eavesdropper state at time slot $l$ as
\begin{equation}
    \hat{\bm\chi}[l|l-1]={\bf F}\hat{\bm\chi}[l-1],
\end{equation}
where $\hat{\bm\chi}[l|l-1]$ denotes the predicted estimate of 
${\bm\chi}[l]$ based on the information available up to time slot $l-1$. 
The corresponding prediction covariance matrix is given by
\begin{equation}
    {\bf \Sigma}[l|l-1]
    =
    {\bf F}{\bf \Sigma}[l-1]{\bf F}^T
    +
    \mathbf{Q}_{{\bf n}_{\bm\chi}}.
\end{equation}}

\subsection{Measurement Model}
\label{sec:st_model}

As illustrated in Fig.~\ref{fig:protocol}, the transmit-signal design and communication stages are completed before sensing measurement collection within each time slot. The SRs then receive the corresponding echo signals and extract the measurements required for EKF-based trajectory tracking.

Since the sensing signals transmitted by different APs are mutually orthogonal, the echo returns associated with different transmitters can be separated at each SR via matched filtering (MF)~\cite{gogineni2011target,haimovich2008mimo,wang2020target,li2021signal,li2008mimo}. Based on the separated echo corresponding to the AP--SR pair $(m,v)$, the measurement vector available at SR $v$ is constructed as
\[
{\bf z}_{m,v}[l]
=
\left[
\hat{\tau}_{m,v}[l],
\hat{\mu}_{m,v}[l],
\sin\hat{\vartheta}_{m,v}[l]
\right]^T,
\]
where $\hat{\tau}_{m,v}[l]$, $\hat{\mu}_{m,v}[l]$, and $\hat{\vartheta}_{m,v}[l]$ denote the noisy estimates of the propagation delay, Doppler frequency, and AoA, respectively. The AoA is represented by its sine to match the angular parameterization adopted in the array response.
Since these measurement parameters depend on the eavesdropper state, the measurement vector is expressed as the nonlinear measurement model
\begin{equation}
{\bf z}_{m,v}[l]
=
{\bf g}_{m,v}({\bm\chi}[l])
+
{\bf n}_{{\bf z}_{m,v}}[l],
\label{eq:meas_model}
\end{equation}
where ${\bf g}_{m,v}:\mathbb{R}^{4}\rightarrow\mathbb{R}^{3}$ denotes the nonlinear measurement function and
${\bf n}_{{\bf z}_{m,v}}[l]\sim\mathcal{N}({\bf0},{\bf R}_{m,v}[l])$
is the corresponding measurement noise.

The covariance matrix ${\bf R}_{m,v}[l]$ characterizes the estimation uncertainty of the delay, Doppler, and AoA measurements. Following standard radar estimation models, its entries are assumed to be inversely proportional to the output signal-to-noise ratio (SNR) of the matched filter, which is given by
\begin{equation}
\mathrm{SNR}_{m,v}[l]
=
\frac{
G_{\rm MF}N_tN_r\beta_0\rho_0
{\bf a}_t^H(\theta_{m,e}[l])
{\bf S}_{m}[l]
{\bf a}_t(\theta_{m,e}[l])
}{
d_{m,e}^2[l]d_{v,e}^2[l]\sigma_{\rm sen}^2
},
\label{eq:senSNR}
\end{equation}
where $G_{\rm MF}$ denotes the matched-filter gain determined by the number of transmit symbols within each time slot, and $\sigma_{\rm sen}^2$ denotes the variance of the equivalent noise at the MF output. Equation~\eqref{eq:senSNR} follows from the unit-norm property of the receive steering vector, i.e.,
${\bf a}_r^H(\theta_{v,e}[l]){\bf a}_r(\theta_{v,e}[l])=1$.

Since the measurement model in \eqref{eq:meas_model} is nonlinear, the EKF employs its first-order linearization around the predicted state estimate. Accordingly, the corresponding Jacobian matrix is given by
\begin{equation}
{\bf G}_{m,v}[l]
=
\left.
\frac{\partial{\bf g}_{m,v}({\bm\chi})}
{\partial{\bm\chi}}
\right|_{{\bm\chi}=\hat{\bm\chi}[l|l-1]}.
\label{eq:jacob}
\end{equation}
The detailed expressions of ${\bf g}_{m,v}(\cdot)$, ${\bf G}_{m,v}[l]$, and ${\bf R}_{m,v}[l]$ are provided in Appendix~\ref{app:EKF}.

\subsection{Measurement Fusion and EKF Update}
The local measurement models derived above are forwarded from the SRs to the CPU through the backhaul links. The CPU then fuses the measurements collected from all AP--SR pairs into a global measurement model for the EKF update~\cite{sijs2008overview}. Specifically, the stacked measurement vector, nonlinear measurement function, Jacobian matrix, and measurement-noise covariance matrix are respectively expressed as
\begin{equation}
{\bf z}[l]
=
\begin{bmatrix}
{\bf z}_{1}^T[l],
{\bf z}_{2}^T[l],
\dots,
{\bf z}_{M}^T[l]
\end{bmatrix}^{T},
\end{equation}
\begin{equation}
{\bf g}({\bm\chi}[l])
=
\begin{bmatrix}
{\bf g}_{1}^T({\bm\chi}[l]),
{\bf g}_{2}^T({\bm\chi}[l]),
\dots,
{\bf g}_{M}^T({\bm\chi}[l])
\end{bmatrix}^{T},
\end{equation}
\begin{equation}
{\bf G}[l]
=
\begin{bmatrix}
{\bf G}_{1}^T[l],
{\bf G}_{2}^T[l],
\dots,
{\bf G}_{M}^T[l]
\end{bmatrix}^{T},
\end{equation}
\begin{equation}
{\bf R}[l]
=
\operatorname{blkdiag}
\left(
{\bf R}_{1}[l],
{\bf R}_{2}[l],
\dots,
{\bf R}_{M}[l]
\right),
\end{equation}
where the AP-wise stacked quantities are defined as
\begin{equation}
\begin{aligned}
{\bf z}_{m}[l]
&=
\begin{bmatrix}
{\bf z}_{m,1}^T[l],
\dots,
{\bf z}_{m,V}^T[l]
\end{bmatrix}^{T},\\
{\bf g}_{m}({\bm\chi}[l])
&=
\begin{bmatrix}
{\bf g}_{m,1}^T({\bm\chi}[l]),
\dots,
{\bf g}_{m,V}^T({\bm\chi}[l])
\end{bmatrix}^{T},\\
{\bf G}_{m}[l]
&=
\begin{bmatrix}
{\bf G}_{m,1}^T[l],
\dots,
{\bf G}_{m,V}^T[l]
\end{bmatrix}^{T},\\
{\bf R}_{m}[l]
&=
\operatorname{blkdiag}
\left(
{\bf R}_{m,1}[l],
\dots,
{\bf R}_{m,V}[l]
\right).
\end{aligned}
\end{equation}

Using the global measurement model above together with the predicted state
$\hat{\bm\chi}[l|l-1]$
and prediction covariance
${\bf\Sigma}[l|l-1]$,
the EKF first computes the Kalman gain as
\begin{equation}
{\bf K}[l]
=
{\bf\Sigma}[l|l-1]
{\bf G}^{T}[l]
\left(
{\bf R}[l]
+
{\bf G}[l]
{\bf\Sigma}[l|l-1]
{\bf G}^{T}[l]
\right)^{-1}.
\label{eq:KG}
\end{equation}

The predicted state is then corrected using the innovation, defined as the difference between the aggregated measurements and their predicted values:
\begin{equation}
\hat{\bm\chi}[l]
=
\hat{\bm\chi}[l|l-1]
+
{\bf K}[l]
\Big(
{\bf z}[l]
-
{\bf g}
(
\hat{\bm\chi}[l|l-1]
)
\Big).
\end{equation}

The corresponding posterior error covariance is updated as
\begin{equation}
{\bf\Sigma}[l]
=
\left(
{\bf I}_4
-
{\bf K}[l]
{\bf G}[l]
\right)
{\bf\Sigma}[l|l-1].
\label{eq:Sigma}
\end{equation}

Substituting \eqref{eq:KG} into \eqref{eq:Sigma} yields
\begin{equation}
{\bf\Sigma}^{-1}[l]
=
{\bf\Sigma}^{-1}[l|l-1]
+
{\bf G}^{T}[l]
{\bf R}^{-1}[l]
{\bf G}[l].
\label{eq:sigmal}
\end{equation}

The posterior covariance matrix quantifies the remaining uncertainty after incorporating all sensing measurements collected during the current time slot. Therefore, its trace,
$\mathrm{Tr}({\bf\Sigma}[l])$,
which corresponds to the posterior MSE of the eavesdropper state estimate, is adopted as the sensing-performance metric throughout this paper.

\section{Uncertainty-Aware Secure Communication}
\label{sec:Uncertainty}
Building on the EKF-based tracking framework presented in Section~\ref{sec:tracking}, this section develops the uncertainty-aware secure communication design. We first construct the nominal eavesdropper channel from the predicted eavesdropper position and derive a position-induced channel uncertainty model. Based on this model, we formulate the robust joint beamforming and sensing covariance optimization problem and subsequently develop a tractable solution algorithm.

\subsection{Position-Induced Channel Uncertainty}

The EKF described in Section~\ref{sec:tracking} provides only a predicted eavesdropper position rather than its exact {position}. Consequently, the transmit design must explicitly account for the resulting channel uncertainty. Under the proposed framework, the transmit signal for time slot $l$ is designed based on the predicted position $\hat{\bf q}_e[l|l-1]$ before the sensing measurements collected in the same slot are used for the EKF update. Accordingly, the eavesdropper channel from AP $m$ is decomposed as
\begin{equation}
{\bf h}_{m,e}[l]
=
\hat{\bf h}_{m,e}[l]
+
\Delta{\bf h}_{m,e}[l],
\end{equation}
where $\hat{\bf h}_{m,e}[l]$ denotes the nominal channel constructed from the predicted position $\hat{\bf q}_e[l|l-1]$, and $\Delta{\bf h}_{m,e}[l]$ represents the channel error caused by the position prediction error.

Since the LoS channel in \eqref{eq:ch_exp} is completely determined by the eavesdropper position through the propagation distance and AoD, the channel error is written as
\begin{align}
\Delta{\bf h}_{m,e}[l]
&=
{\bf h}_{m,e}({\bf q}_e[l])
-
{\bf h}_{m,e}(\hat{\bf q}_e[l|l-1]),
\end{align}
where ${\bf h}_{m,e}(\cdot)$ denotes the channel expression in \eqref{eq:ch_exp} parameterized by the eavesdropper position. Since the adopted geometric channel model is continuously differentiable with respect to the eavesdropper position, the channel perturbation can be approximated by the first-order Taylor expansion around the predicted position, yielding
\begin{equation}
\Delta{\bf h}_{m,e}[l]
\approx
{\bf J}_m[l]
\Delta{\bf q}_e[l],
\end{equation}
where
\begin{equation}
\Delta{\bf q}_e[l]
\triangleq
{\bf q}_e[l]
-
\hat{\bf q}_e[l|l-1]
=
\begin{bmatrix}
\Delta x_e[l]\\
\Delta y_e[l]
\end{bmatrix},
\end{equation}
and
\begin{equation}
{\bf J}_m[l]
=
\left.
\frac{\partial{\bf h}_{m,e}({\bf q}_e)}
{\partial{\bf q}_e^T}
\right|_{{\bf q}_e=\hat{\bf q}_e[l|l-1]}.
\end{equation}
Stacking the channel vectors associated with all APs gives
\begin{equation}
\Delta{\bf h}_e[l]
=
\begin{bmatrix}
\Delta{\bf h}_{1,e}^T[l]
&
\cdots
&
\Delta{\bf h}_{M,e}^T[l]
\end{bmatrix}^{T}
\approx
{\bf J}[l]
\Delta{\bf q}_e[l],
\label{eq:ch_approx}
\end{equation}
where
\begin{equation}
{\bf J}[l]
=
\begin{bmatrix}
{\bf J}_1^T[l]
&
\cdots
&
{\bf J}_M^T[l]
\end{bmatrix}^{T}.
\end{equation}
Accordingly, the aggregate eavesdropper channel is approximated as
\begin{equation}
{\bf h}_e[l]
=
\hat{\bf h}_e[l]
+
{\bf J}[l]
\Delta{\bf q}_e[l].
\label{eq:quad_form}
\end{equation}

Since the true eavesdropper position is unavailable, the prediction error $\Delta{\bf q}_e[l]$ cannot be directly observed. Instead, we construct a deterministic uncertainty set from the prediction covariance ${\bf\Sigma}[l|l-1]$. Specifically, the component-wise uncertainty intervals are defined as
\begin{equation}
\Omega_{\Delta x_e[l]}
=
\left\{
\Delta x_e[l]
\;\middle|\;
|\Delta x_e[l]|
\le
\eta\sigma_{\Delta x_e[l]}
\right\},
\end{equation}
and
\begin{equation}
\Omega_{\Delta y_e[l]}
=
\left\{
\Delta y_e[l]
\;\middle|\;
|\Delta y_e[l]|
\le
\eta\sigma_{\Delta y_e[l]}
\right\},
\end{equation}
where
$\sigma_{\Delta x_e[l]}
=
\sqrt{[{\bf\Sigma}[l|l-1]]_{1,1}}$
and
$\sigma_{\Delta y_e[l]}
=
\sqrt{[{\bf\Sigma}[l|l-1]]_{2,2}}$ are the predicted standard deviations of the position errors. The parameter $\eta>0$ specifies the confidence level of the uncertainty region, with a larger $\eta$ yielding a more conservative robust design. 

To facilitate robust beamforming, the rectangular uncertainty box is approximated by the following ellipsoidal uncertainty region:
\begin{equation}
\Omega_{\Delta{\bf q}_e[l]}
=
\left\{
\Delta{\bf q}_e[l]
\;\middle|\;
\Delta{\bf q}_e^T[l]
{\bm\Psi}[l]
\Delta{\bf q}_e[l]
\le
1
\right\},
\label{eq:uncertainty_region}
\end{equation}
where
\begin{equation}
{\bm\Psi}[l]
=
\operatorname{diag}
\left(
\frac{1}{(\eta\sqrt2\sigma_{\Delta x_e[l]})^2},
\frac{1}{(\eta\sqrt2\sigma_{\Delta y_e[l]})^2}
\right).
\end{equation}
The factor $\sqrt2$ guarantees that the ellipsoidal region contains the rectangular uncertainty box.

Based on the above uncertainty model, we next define the secrecy performance metric. Let ${\bf w}[l]=\{{\bf w}_1[l],\ldots,{\bf w}_K[l]\}$ denote the communication beamforming vectors. The achievable rate of legitimate user $k$ is given by 
\begin{multline}
R_{k}({\bf w}[l],{\bf S}[l])\\
=
\log_2\left(
1+
\frac{
|\mathbf{h}_{k}^H \mathbf{w}_{k}[l] |^2
}{
\sum_{k' \in \mathcal{K}\setminus\{k\}} |\mathbf{h}_{k}^H \mathbf{w}_{k'}[l]|^2
+
{\bf h}_k^H{\bf S}[l]{\bf h}_k
+
\sigma_k^2
}
\right).
\label{eq:user_rate}
\end{multline}
For a conservative secrecy analysis, we assume that the eavesdropper can perfectly cancel the interference caused by the communication signals intended for the other users~\cite{strong_eve}. Under this strong-eavesdropper model, the eavesdropper is impaired only by the sensing interference and receiver noise. Accordingly, the achievable rate at the eavesdropper for decoding the message intended for user $k$ at time slot $l$ is given by
\begin{equation}
\begin{aligned}
R_{e,k}(&{\bf w}_k[l],{\bf S}[l],\Delta{\bf q}_e[l])\!=\!
\log_2\!\left(
1+
\frac{
|{\bf h}_e^H[l]{\bf w}_k[l]|^2
}{
{\bf h}_e^H[l]{\bf S}[l]{\bf h}_e[l]
+
\sigma_e^2
}
\right),
\end{aligned}
\label{eq:eve_rate}
\end{equation}
where ${\bf h}_e[l]$ is given by \eqref{eq:quad_form} with
$\Delta{\bf q}_e[l]\in\Omega_{\Delta{\bf q}_e[l]}$.

Based on the uncertainty region in \eqref{eq:uncertainty_region}, we adopt the worst-case secrecy rate as the secrecy performance metric~\cite{wcssr}. Specifically, the worst-case secrecy rate of user $k$ at time slot $l$ is defined as
\begin{equation}
\begin{aligned}
&R_{{\rm sec},k}({\bf w}[l],{\bf S}[l])=
\\&
\left[
R_k({\bf w}[l],{\bf S}[l])
-
\max_{\Delta{\bf q}_e[l]\in\Omega_{\Delta{\bf q}_e[l]}}
R_{e,k}
({\bf w}_k[l],{\bf S}[l],\Delta{\bf q}_e[l])
\right]^+,
\end{aligned}
\label{eq:sec_Rate}
\end{equation}
where $[x]^+\triangleq\max(x,0)$.

\subsection{Robust Joint Beamforming and Sensing Design}
\label{sec:opt}
Building on the position-induced channel uncertainty model developed in the previous subsection, we now optimize the transmit strategy for secure communication. Specifically, the CPU jointly designs the communication beamforming vectors and sensing covariance matrices to maximize the worst-case sum secrecy rate while satisfying the per-AP transmit-power constraints and a sensing accuracy requirement. We first formulate the robust optimization problem and then develop an efficient algorithm for its solution.

\subsubsection{Problem Formulation}
At each time slot, the CPU jointly optimizes the communication beamforming vectors and sensing covariance matrices based on the predicted eavesdropper state. The resulting robust transmit design problem is formulated as
\begin{subequations}\label{eq:P1}
\begin{align}
\max_{\mathbf{w}[l],{\bf S}[l]} \ 
& \sum_{k\in\mathcal{K}} R_{{\rm sec},k}({\bf w}[l],{\bf S}[l])
\label{eq:P1a}\\
\text{s.t.}\ 
& \operatorname{Tr}({\bf S}_{m}[l])
+\sum_{k\in\mathcal{K}}\|{\bf w}_{m,k}[l]\|^2
\leq P_{\max},\quad \forall m,
\label{eq:P1b}\\
& \operatorname{Tr}({\bf \Sigma}[l])\leq\Gamma_{\rm th},
\label{eq:P1c}\\
& {\bf S}_m[l]\succeq{\bf 0},\quad \forall m.
\label{eq:P1d}
\end{align}
\end{subequations}

Constraint \eqref{eq:P1b} imposes the per-AP transmit-power budget, while \eqref{eq:P1c} guarantees the required sensing accuracy by limiting the EKF posterior covariance. Constraint \eqref{eq:P1d} enforces the positive semidefiniteness of the sensing covariance matrices. Here, the posterior covariance ${\bf\Sigma}[l]$ depends on both the predicted eavesdropper state and the sensing quality induced by the transmit design.
Problem~\eqref{eq:P1} is highly non-convex for two reasons. First, the secrecy-rate objective involves the worst-case eavesdropping rate over the position uncertainty region. Second, the sensing constraint is non-convex because the EKF posterior covariance depends nonlinearly on the transmit design variables through the sensing SNR.

For notational simplicity, the time-slot index $l$ is omitted hereafter. To obtain a more tractable formulation, we first remove the operator $[\cdot]^+$ in \eqref{eq:sec_Rate}, which is valid because the optimal solution always yields a non-negative secrecy rate~\cite[Lemma~3]{taghizadeh2019secrecy}. We then introduce auxiliary variables $\gamma_k\ge0$, $\forall k\in\mathcal K$, to upper-bound the worst-case eavesdropping signal-to-interference-plus-noise ratios (SINRs). Since the achievable rate at the eavesdropper is a monotonically increasing function of the SINR, these auxiliary variables equivalently upper-bound the corresponding worst-case eavesdropping rates. Accordingly, Problem~\eqref{eq:P1} can be reformulated as
\begin{subequations}\label{eq:P2}
\begin{align}
\max_{\mathbf{w},{\bf S},{\bm\gamma}} \ 
& \sum_{k\in\mathcal{K}}
\left(
R_k({\bf w},{\bf S})
-
\log_2(1+\gamma_k)
\right)
\label{eq:P2a}\\
\text{s.t.}\ 
& \max_{\Delta{\bf q}_e\in\Omega_{\Delta{\bf q}_e}}
\frac{
|{\bf h}_e^H{\bf w}_k|^2
}{
{\bf h}_e^H{\bf S}{\bf h}_e
+
\sigma_e^2
}
\le\gamma_k,\quad\forall k,
\label{eq:P2b}\\
& \gamma_k\ge0,\quad\forall k,
\label{eq:P2c}\\
& \eqref{eq:P1b},~\eqref{eq:P1c},~\eqref{eq:P1d}, \notag
\end{align}
\end{subequations}
where ${\bm\gamma}\triangleq[\gamma_1,\ldots,\gamma_K]^T$.
Although Problem~\eqref{eq:P2} has a more convenient objective, it remains non-convex due to the coupled transmit variables and worst-case SINR constraints. To address this challenge, we adopt a two-block AO framework~\cite{razaviyayn2013unified}, in which the transmit-variable block $\{{\bf w},{\bf S}\}$ and the auxiliary-variable block ${\bm\gamma}$ are optimized alternately, with one block fixed while the other is optimized. As shown in the following subsections, each resulting subproblem can be transformed into a convex semidefinite program by appropriately applying semidefinite relaxation (SDR)~\cite{luo2010semidefinite}, the S-procedure~\cite{boyd2004convex}, and the Schur complement~\cite{zhang2006schur}.

\subsubsection{Transmit Variable Optimization}
\label{sec:sub1}
For fixed ${\bm\gamma}$, the terms $-\log_2(1+\gamma_k)$ in the objective of Problem~\eqref{eq:P2} become constants independent of the transmit variables $\{{\bf w},{\bf S}\}$. Accordingly, the corresponding AO subproblem is given by
\begin{subequations}\label{eq:Pw0}
\begin{align}
\max_{\mathbf{w},{\bf S}} \ 
& \sum_{k\in\mathcal{K}} R_k({\bf w},{\bf S})
\label{eq:Pw0a}\\
\text{s.t.}\ 
&\eqref{eq:P1b},~\eqref{eq:P1c},~\eqref{eq:P1d},~\eqref{eq:P2b}. \notag
\end{align}
\end{subequations}

Problem~\eqref{eq:Pw0} is non-convex due to the non-concave achievable-rate objective and the worst-case eavesdropping SINR constraint. To facilitate a tractable reformulation, we introduce the lifted beamforming matrices
\[
{\bf W}_k \triangleq {\bf w}_k{\bf w}_k^H,\qquad
\forall k\in\mathcal K,
\]
and denote
\[
{\bf W}\triangleq\{{\bf W}_1,\ldots,{\bf W}_K\}.
\]
For each AP $m$, ${\bf W}_{m,k}\in\mathbb{C}^{N_t\times N_t}$ denotes the $m$-th principal diagonal block of ${\bf W}_k$. By construction, each lifted matrix satisfies
${\bf W}_k\succeq{\bf0}$
and
$\operatorname{rank}({\bf W}_k)=1$.
Accordingly, Problem~\eqref{eq:Pw0} can be equivalently reformulated as
\begin{subequations}\label{eq:Pw1}
\begin{align}
\max_{\mathbf{W},{\bf S}} \ 
& \sum_{k\in\mathcal{K}} R_k({\bf W},{\bf S})
\label{eq:Pw1a}\\
\text{s.t.}\ 
&
\max_{\Delta{\bf q}_e\in\Omega_{\Delta{\bf q}_e}}
\frac{
{\bf h}_e^H{\bf W}_k{\bf h}_e
}{
{\bf h}_e^H{\bf S}{\bf h}_e+\sigma_e^2
}
\le\gamma_k,
\quad\forall k,
\label{eq:Pw1b}
\\
&
\operatorname{Tr}({\bf S}_m)
+
\sum_{k\in\mathcal K}
\operatorname{Tr}({\bf W}_{m,k})
\le
P_{\max},
\quad\forall m,
\label{eq:Pw1c}
\\
&
{\bf W}_k\succeq{\bf0},
\quad\forall k,
\label{eq:Pw1d}
\\
&
\operatorname{rank}({\bf W}_k)=1,
\quad\forall k,
\label{eq:Pw1e}
\\
&
\eqref{eq:P1c},~\eqref{eq:P1d}, \notag
\end{align}
\end{subequations}
where
\begin{equation}
\begin{aligned}
&R_k({\bf W},{\bf S})
=\\
&\log_2\!\left(
1+
\frac{
{\bf h}_k^H{\bf W}_k{\bf h}_k
}{
\sum_{k'\in\mathcal K\setminus\{k\}}
{\bf h}_k^H{\bf W}_{k'}{\bf h}_k
+
{\bf h}_k^H{\bf S}{\bf h}_k
+
\sigma_k^2
}
\right).
\end{aligned}
\end{equation}
Problem~\eqref{eq:Pw1} remains non-convex due to the rank-one constraints in \eqref{eq:Pw1e}. We therefore first apply SDR by removing the rank-one constraints. Even after SDR, the resulting problem remains challenging for three reasons. First, the achievable-rate objective is non-concave. Second, the worst-case eavesdropping SINR constraint in \eqref{eq:Pw1b} is a semi-infinite constraint over the continuous position uncertainty region. Third, the sensing constraint in \eqref{eq:P1c} depends nonlinearly on the EKF posterior covariance. These three issues are addressed sequentially.

First, we address the non-concavity of the achievable-rate objective. To this end, we rewrite $R_k({\bf W},{\bf S})$ in the following difference-of-convex form:
\begin{equation}
\begin{aligned}
R_k({\bf W},{\bf S})
&=
\log_2\!\left(
{\bf h}_k^H{\bf W}_k{\bf h}_k
+
T_k({\bf W},{\bf S})
\right)
\\
&\quad
-
\log_2\!\left(
T_k({\bf W},{\bf S})
\right),
\end{aligned}
\label{eq:DC}
\end{equation}
where
\begin{equation}
T_k({\bf W},{\bf S})
\triangleq
\sum_{k'\in\mathcal K\setminus\{k\}}
{\bf h}_k^H{\bf W}_{k'}{\bf h}_k
+
{\bf h}_k^H{\bf S}{\bf h}_k
+
\sigma_k^2.
\end{equation}

Since the second term in \eqref{eq:DC} is convex, it can be lower-bounded by its first-order Taylor expansion at the current feasible point~\cite{razaviyayn2013unified}. Denoting this affine lower bound by $\phi_k({\bf W},{\bf S})$, we obtain the following concave lower bound on $R_k({\bf W},{\bf S})$:
\begin{equation}
\underline{R}_k({\bf W},{\bf S})
=
\log_2\!\left(
{\bf h}_k^H{\bf W}_k{\bf h}_k
+
T_k({\bf W},{\bf S})
\right)
+
\phi_k({\bf W},{\bf S}),
\label{eq:LBP3}
\end{equation}
where
\begin{align}
\phi_k({\bf W},{\bf S})
&\triangleq
-\log_2(\tilde T_k)
-
\operatorname{Tr}
\!\left(
\frac{{\bf h}_k{\bf h}_k^H}
{\tilde T_k\ln2}
({\bf S}-\tilde{\bf S})
\right)
\\
&-
\sum_{k'\in\mathcal K\setminus\{k\}}
\operatorname{Tr}
\!\left(
\frac{{\bf h}_k{\bf h}_k^H}
{\tilde T_k\ln2}
({\bf W}_{k'}-\tilde{\bf W}_{k'})
\right). 
\end{align}
Here, $\{\tilde{\bf W},\tilde{\bf S}\}$ denotes the current feasible point and $\tilde T_k \triangleq T_k(\tilde{\bf W},\tilde{\bf S})$.

Since
$\underline{R}_k({\bf W},{\bf S})$
is concave and satisfies
\[
\underline{R}_k({\bf W},{\bf S})
\le
R_k({\bf W},{\bf S}),
\]
we replace
$R_k({\bf W},{\bf S})$
in the objective of Problem~\eqref{eq:Pw1}
with
$\underline{R}_k({\bf W},{\bf S})$
to obtain a concave approximation.

Next, we reformulate the worst-case eavesdropping SINR constraint in \eqref{eq:Pw1b}. By substituting the channel uncertainty model
${\bf h}_e=\hat{\bf h}_e+{\bf J}\Delta{\bf q}_e$
into \eqref{eq:Pw1b}, the semi-infinite constraint can be expressed as
\begin{multline}
(\hat{\bf h}_e+{\bf J}\Delta{\bf q}_e)^H
({\bf W}_k-\gamma_k{\bf S})
(\hat{\bf h}_e+{\bf J}\Delta{\bf q}_e)
-\gamma_k\sigma_e^2
\le0,\\
\forall k,\;
\forall\Delta{\bf q}_e\in\Omega_{\Delta{\bf q}_e}.
\label{eq:uncer}
\end{multline}

Using the uncertainty region in \eqref{eq:uncertainty_region}, \eqref{eq:uncer} can be equivalently rewritten as an implication between two quadratic inequalities:
\begin{multline}
\Delta{\bf q}_e^T{\bm\Psi}\Delta{\bf q}_e-1\le0
\\
\Rightarrow
\Delta{\bf q}_e^H{\bf A}_k\Delta{\bf q}_e
+
{2\Re\!\left({\bf b}_k^H\Delta{\bf q}_e\right)}
+
C_k
\le0,
\quad
\forall k,
\label{eq:sform_final}
\end{multline}
where
\begin{align}
{\bf A}_k
&=
{\bf J}^H({\bf W}_k-\gamma_k{\bf S}){\bf J},\\
{{\bf b}_k}
&=
{\bf J}^H({\bf W}_k-\gamma_k{\bf S})\hat{\bf h}_e,
\end{align}
and
\begin{align}
C_k
=
\hat{\bf h}_e^H({\bf W}_k-\gamma_k{\bf S})\hat{\bf h}_e
-\gamma_k\sigma_e^2.    
\end{align}

The above implication between quadratic inequalities can be equivalently transformed into a linear matrix inequality (LMI) via the S-procedure~\cite{boyd2004convex}.
\begin{theorem}[S-procedure]
\label{theom1}
Let
$
f_i({\bf x})
=
{\bf x}^H{\bf A}_i{\bf x}
+
2\Re\{{\bf b}_i^H{\bf x}\}
+
c_i,
$
where
${\bf A}_i\in\mathbb H^n$,
${\bf b}_i\in\mathbb C^n$,
and
$c_i\in\mathbb R$,
for
$i\in\{1,2\}$.
Then,
$f_1({\bf x})\le0\Rightarrow f_2({\bf x})\le0$
holds if and only if there exists a scalar
$\kappa\ge0$
such that
\begin{equation}
\begin{bmatrix}
{\bf A}_2 & {\bf b}_2\\
{\bf b}_2^H & c_2
\end{bmatrix}
\preceq
\kappa
\begin{bmatrix}
{\bf A}_1 & {\bf b}_1\\
{\bf b}_1^H & c_1
\end{bmatrix}.
\end{equation}
\end{theorem}

Applying Theorem~\ref{theom1} to \eqref{eq:sform_final} yields the following LMI:
\begin{equation}
\begin{bmatrix}
\kappa_k{\bm\Psi}-{\bf A}_k
&
{-{\bf b}_k}
\\
{-{\bf b}_k^H}
&
-\kappa_k-C_k
\end{bmatrix}
\succeq{\bf0},
\quad
\kappa_k\ge0,
\quad
\forall k.
\label{eq:M_k}
\end{equation}

Finally, we reformulate the sensing MSE constraint in \eqref{eq:P1c}. Since
$\operatorname{Tr}({\bf\Sigma})=\sum_{i=1}^{4}[{\bf\Sigma}]_{i,i}$, we introduce auxiliary variables $c_i\ge0$, $i\in\mathcal I\triangleq\{1,2,3,4\}$, satisfying
\[
[{\bf\Sigma}]_{i,i}\le c_i,\qquad \forall i.
\]
Using the Schur complement~\cite{zhang2006schur}, the sensing constraint can be equivalently transformed into the following LMI:
\begin{equation}
\begin{gathered}
\begin{bmatrix}
{\bf\Sigma}^{-1} & {\bf e}_i\\
{\bf e}_i^T & c_i
\end{bmatrix}
\succeq{\bf0},
\quad
c_i\ge0,\quad\forall i,\\
\sum_{i=1}^4 c_i
\le
\Gamma_{\rm th},
\end{gathered}
\label{eq:EKFLMI}
\end{equation}
where ${\bf e}_i\in\mathbb R^{4\times1}$ denotes the $i$-th column of ${\bf I}_4$.

Combining the above reformulations yields the following convex semidefinite program (SDP):
\begin{subequations}\label{eq:Pw2}
\begin{align}
\max_{\mathbf{W},{\bf S},{\bm \kappa},{\bf c}} \ 
& \sum_{k\in\mathcal{K}}
\underline{R}_k({\bf W},{\bf S})
\label{eq:Pw2a}\\
\text{s.t.}\ 
&\eqref{eq:P1d},~\eqref{eq:Pw1c},~\eqref{eq:Pw1d},~\eqref{eq:M_k},~\eqref{eq:EKFLMI}, \notag
\end{align}
\end{subequations}
where
${\bm\kappa}\triangleq[\kappa_1,\ldots,\kappa_K]^T$
and
${\bf c}\triangleq[c_1,c_2,c_3,c_4]^T$. Problem~\eqref{eq:Pw2} is a convex SDP and can therefore be efficiently solved using CVX~\cite{grant2014cvx}. The obtained solution is expressed in terms of the lifted beamforming matrices $\{{\bf W}_k\}$.

If the optimal solution satisfies the rank-one constraints, the beamforming vectors are directly recovered from $\{{\bf W}_k\}$. Otherwise, we recover the beamforming vectors from the dominant eigenvectors of the relaxed solutions instead of applying Gaussian randomization. This approach has substantially lower computational complexity than Gaussian randomization while providing an effective rank-one approximation for the considered transmit beamforming design. Accordingly, an approximate beamforming vector for user $k$ is constructed via eigenvalue decomposition (EVD) as
\[
{\bf w}_k
=
\alpha_k
\sqrt{\lambda_{\max}({\bf W}_k)}
{\bf v}_{\max}({\bf W}_k),
\]
where
$\lambda_{\max}({\bf W}_k)$
and
${\bf v}_{\max}({\bf W}_k)$
denote the largest eigenvalue of
${\bf W}_k$
and its corresponding eigenvector, respectively, and
$\alpha_k\in(0,1]$
is chosen {to satisfy the per-AP power constraints.}

\subsubsection{Auxiliary Variable Optimization}

For fixed transmit variables $\{{\bf W},{\bf S}\}$, the achievable-rate terms become constants with respect to ${\bm\gamma}$. Accordingly, the corresponding AO subproblem is given by
\begin{subequations}\label{eq:Pg1}
\begin{align}
\max_{{\bm\gamma}}
\quad
&
\sum_{k\in\mathcal K}
-\log_2(1+\gamma_k)
\label{eq:Pg1a}
\\
\text{s.t.}\quad
&
\eqref{eq:P2c},~\eqref{eq:Pw1b}. \notag
\end{align}
\end{subequations}

Problem~\eqref{eq:Pg1} is non-convex because the objective is a maximization of the convex function $-\log_2(1+\gamma_k)$. Moreover, the worst-case eavesdropping SINR constraint is represented by the semi-infinite constraint in \eqref{eq:Pw1b}. Using the equivalent LMI reformulation in \eqref{eq:M_k}, the remaining non-convexity lies only in the objective function.

To obtain a tractable approximation, we replace $-\log_2(1+\gamma_k)$ by its first-order Taylor lower bound at the current feasible point $\tilde{\gamma}_k$~\cite{razaviyayn2013unified}, i.e.,
\begin{equation}
\psi_k(\gamma_k)
\triangleq
-\log_2(1+\tilde{\gamma}_k)
-
\frac{\gamma_k-\tilde{\gamma}_k}
{(1+\tilde{\gamma}_k)\ln2}.
\end{equation}
Since
\[
\psi_k(\gamma_k)
\le
-\log_2(1+\gamma_k),
\]
and $\psi_k(\gamma_k)$ is affine, the resulting surrogate objective is concave with respect to ${\bm\gamma}$.

Combining the above approximation with the LMI constraint in \eqref{eq:M_k} yields the following convex subproblem:
\begin{subequations}\label{eq:Pg2}
\begin{align}
\max_{{\bm\gamma},{\bm\kappa}}
\quad
&
\sum_{k\in\mathcal K}
\psi_k(\gamma_k)
\label{eq:Pg2a}
\\
\text{s.t.}\quad
&
\eqref{eq:P2c},~\eqref{eq:M_k}. \notag
\end{align}
\end{subequations}

Problem~\eqref{eq:Pg2} is convex and can therefore be efficiently solved using CVX~\cite{grant2014cvx}.

\begin{algorithm}[t]
\caption{AO-based algorithm for robust beamforming optimization}
\label{alg:Algorithm1}
\begin{algorithmic}[1]
\STATE {\textbf{Input:} $P_{\max}$, $\Gamma_{\rm th}$, $\epsilon$}
\STATE {Set $p=0$, $\tilde{R}_{{\rm sec}}^{(0)}=0$, and initialize ${\bf w}^{(0)}$, ${\bf S}^{(0)}$, ${\bm\gamma}^{(0)}$}
\WHILE{$p=0$ \OR $|(\tilde{R}_{{\rm sec}}^{(p)}-\tilde{R}_{{\rm sec}}^{(p-1)})/\tilde{R}_{{\rm sec}}^{(p)}| > \epsilon$}
\STATE $p \leftarrow p + 1$
\STATE ${\bm\gamma} \leftarrow {\bm\gamma}^{(p-1)}$
\STATE Solve Problem~\eqref{eq:Pw2} and obtain $\{{\bf W}^{(p)},{\bf S}^{(p)}\}$
\STATE Update ${\bf w}_k^{(p)}\leftarrow \alpha_k\sqrt{\lambda_{\max}\left({\bf W}_k^{(p)}\right)}{\bf v}_{\max}\left({\bf W}_k^{(p)}\right),\forall k$ 
\STATE $\{{\bf w},{\bf S}\}\leftarrow \{{\bf w}^{(p)},{\bf S}^{(p)}\}$
\STATE Solve Problem~\eqref{eq:Pg2} and obtain ${\bm\gamma}^{(p)}$ 
\STATE Calculate $\tilde{R}_{{\rm sec}}^{(p)}$
\ENDWHILE
\STATE \textbf{Output:} $\{{\bf w}^{(p)},{\bf S}^{(p)},{\bm\gamma}^{(p)}\}$
\end{algorithmic}
\end{algorithm}

\subsubsection{Overall Algorithm and Complexity Analysis}
The proposed AO algorithm alternately solves the two convex subproblems developed above. Let
$\{{\bf w}^{(p)},{\bf S}^{(p)},{\bm\gamma}^{(p)}\}$
denote the optimization variables obtained at the $p$-th AO iteration, and let
$\tilde{R}_{\rm sec}^{(p)}$
denote the corresponding objective value of Problem~\eqref{eq:P2}.

Specifically, at the $p$-th AO iteration, Problem~\eqref{eq:Pw2} is first solved for fixed
${\bm\gamma}^{(p-1)}$
to obtain the relaxed solution
$\{{\bf W}^{(p)},{\bf S}^{(p)}\}$.
The corresponding beamforming vectors
$\{{\bf w}^{(p)}\}$
are then recovered via the EVD procedure described above.
Using the updated transmit variables,
Problem~\eqref{eq:Pg2}
is subsequently solved to obtain
${\bm\gamma}^{(p)}$.
The iterations terminate when the relative change in
$\tilde{R}_{\rm sec}^{(p)}$
falls below a prescribed tolerance.
The overall procedure is summarized in Algorithm~\ref{alg:Algorithm1}.

The convergence of Algorithm~\ref{alg:Algorithm1} follows from the successive convex approximation framework~\cite{razaviyayn2013unified}. Specifically, in the transmit-variable update, the achievable-rate objective is replaced by a concave lower bound that is tight at the current iterate. Likewise, in the auxiliary-variable update, each term $-\log_2(1+\gamma_k)$ is replaced by its affine lower bound, which is also tight at the current iterate. Therefore, each AO iteration produces a non-decreasing objective value. Since the objective is upper-bounded under the per-AP transmit-power constraints, the sequence
$\{\tilde{R}_{\rm sec}^{(p)}\}$
is guaranteed to converge.

We next analyze the computational complexity of Algorithm~\ref{alg:Algorithm1}. Both subproblems are solved using the interior-point method implemented in CVX. Following the commonly adopted complexity approximation in~\cite{son2025group}, the computational complexity of each convex subproblem is approximated by
$\mathcal{O}(D^{3.5})$,
where
$D$
denotes the number of dominant scalar optimization variables.
For Problem~\eqref{eq:Pw2}, the dominant optimization variables consist of the
$K$
lifted beamforming matrices
${\bf W}_k\in\mathbb{C}^{MN_t\times MN_t}$
and the block-diagonal sensing covariance matrix
${\bf S}$,
which contains
$M$
blocks of size
$N_t\times N_t$.
Ignoring the auxiliary variables due to their relatively small dimensions, the number of dominant scalar variables is on the order of
\[
K(MN_t)^2+MN_t^2.
\]
Hence, the computational complexity of solving Problem~\eqref{eq:Pw2} is approximately
\[
\mathcal{O}\!\left(
\left(
K(MN_t)^2+MN_t^2
\right)^{3.5}
\right).
\]

In contrast, Problem~\eqref{eq:Pg2} involves only
$K$
auxiliary variables together with the associated LMI multipliers, whose dimensions are much smaller than those of Problem~\eqref{eq:Pw2}. Therefore, its computational complexity is negligible in comparison. Consequently, letting
$I_{\rm iter}$
denote the number of AO iterations required for convergence, the overall computational complexity of Algorithm~\ref{alg:Algorithm1} is approximated as
\[
\mathcal{O}\!\left(
I_{\rm iter}
\left(
K(MN_t)^2+MN_t^2
\right)^{3.5}
\right).
\]

\section{Simulation Results}
\label{sec:simulation}
\begin{figure}
\centering
\includegraphics[width=0.7\linewidth]{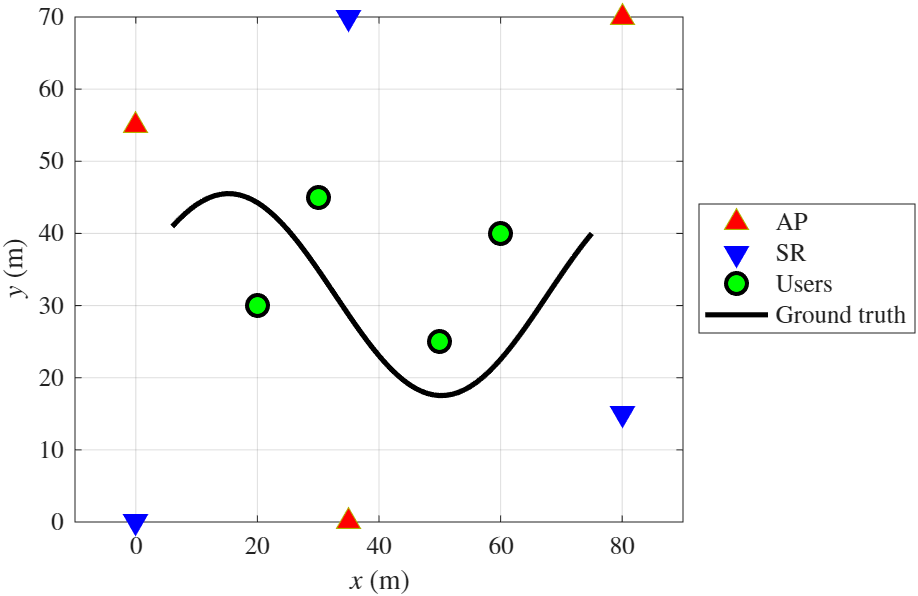}
\caption{{Simulation topology.}}
\label{fig:model}
\end{figure}
This section presents simulation results to validate the effectiveness of the proposed framework.
As illustrated in Fig.~\ref{fig:model}, the considered system consists of $K=4$ legitimate users, $M=3$ APs, and $V=3$ SRs.
Each AP and SR is equipped with an $N$-element ULA, i.e., $N_t=N_r=N$. Unless otherwise specified, we set $N=8$.
The carrier frequency is set to $f_c=28$ GHz.
The channel power gain at a reference distance of $1\,\mathrm{m}$ is given by $\beta_0=(c/(4\pi f_c))^2$~\cite{xu2025sensing}, where $c$ denotes the speed of light.
Unless otherwise specified, the simulation parameters are set as follows:
$P_{\max}=30$ dBm,
$\sigma_k^2=-80$ dBm,
$\sigma_e^2=-80$ dBm,
$\sigma_{\rm sen}^2=-80$ dBm,
$G_{\rm MF}=10^3$,
$\rho_0=0.1$,
$\sigma_{x_e}=0.1$ m,
$\sigma_{y_e}=0.1$ m,
$\sigma_{\dot{x}_e}=0.1$ m/s,
$\sigma_{\dot{y}_e}=0.5$ m/s,
$\Delta t=0.1$ s,
$L=70$,
$\Gamma_{\rm th}=3$, and
$\eta=3$.
The measurement noise scaling coefficients are set to
$a_{\tau}=10^{-6}$,
$a_{\mu}=2\times10^4$, and
$a_{\vartheta}=1$~\cite{zhao2025sensing}.

\begin{figure}
\centering
\includegraphics[width=0.5\linewidth]{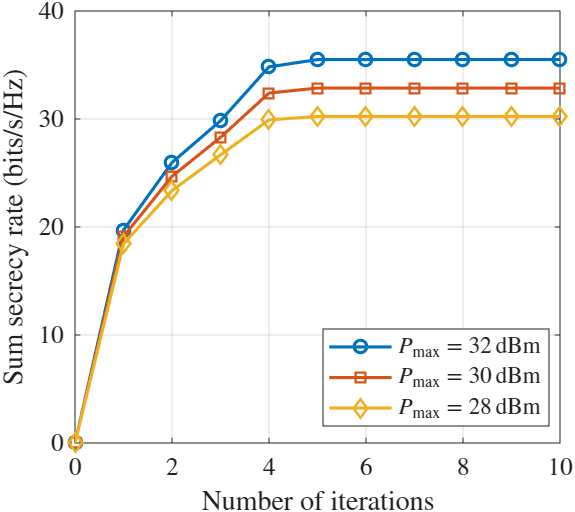}
\caption{{Convergence of the proposed robust transmit design algorithm.}}
\label{fig:converge}
\end{figure}
\begin{figure}[!t]
\centering
\includegraphics[width=0.75\linewidth]{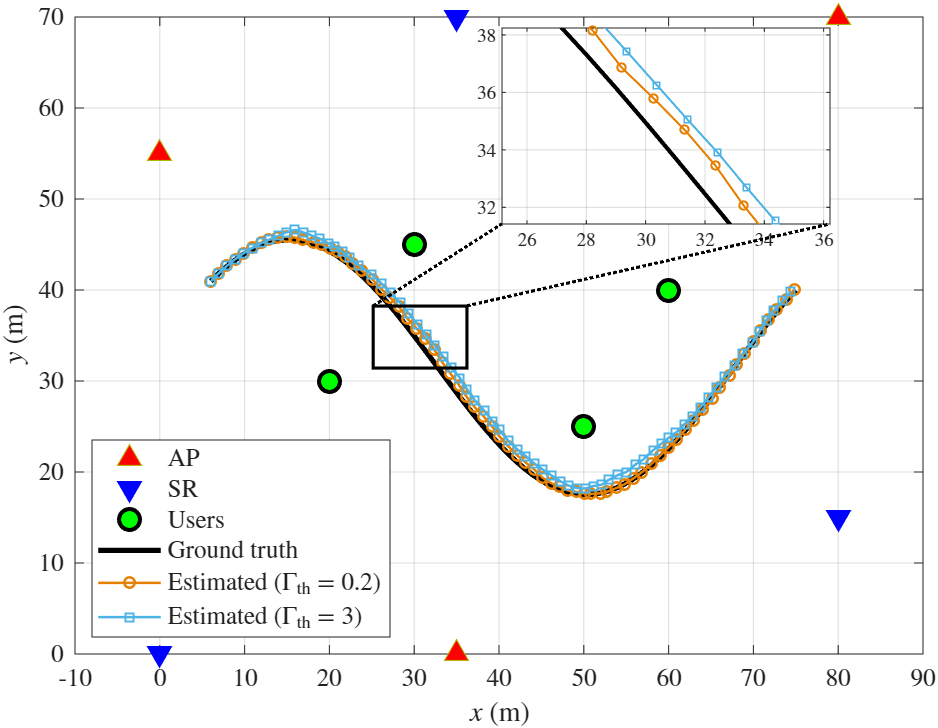}
\caption{Eavesdropper trajectory tracking under different sensing MSE thresholds $\Gamma_{\rm th}$.}
\label{fig:traj_zoom}
\end{figure}

We first examine the convergence behavior of the proposed algorithm.
Fig.~\ref{fig:converge} depicts the objective value of Algorithm~\ref{alg:Algorithm1} versus the iteration number for different per-AP transmit-power budgets.
The objective value monotonically increases for all considered values of $P_{\max}$ and converges within five iterations.
These results demonstrate that the proposed AO-based algorithm converges rapidly and exhibits stable convergence behavior under different transmit-power budgets.

\begin{figure}[!t]
\centering
\includegraphics[width=0.6\linewidth]{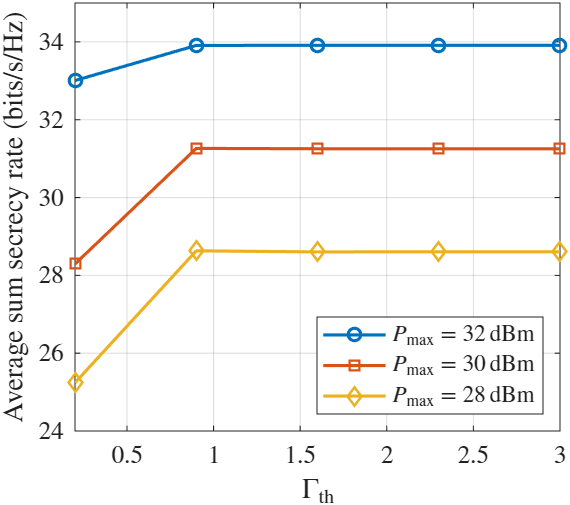}
\caption{Average sum secrecy rate versus the sensing MSE threshold $\Gamma_{\rm th}$ for different values of $P_{\max}$.}
\label{fig:average}
\end{figure}

As shown in Fig.~\ref{fig:traj_zoom}, the estimated trajectory obtained with $\Gamma_{\rm th}=0.2$ follows the ground-truth trajectory more closely than that obtained with $\Gamma_{\rm th}=3$.
This is because a smaller $\Gamma_{\rm th}$ imposes a more stringent sensing MSE requirement, thereby placing greater emphasis on improving the tracking accuracy in the transmit design.
{Fig.~\ref{fig:average} further examines the impact of the sensing MSE requirement on secrecy performance by showing the average sum secrecy rate versus $\Gamma_{\rm th}$.}
Specifically, the average sum secrecy rate increases as $\Gamma_{\rm th}$ increases from $0.2$ to $0.9$ and then remains nearly unchanged for all considered transmit-power levels.
This indicates that an overly stringent sensing requirement sacrifices secrecy performance, whereas a moderately relaxed requirement allows the proposed design to maintain sufficiently accurate tracking while improving the secrecy rate.
Once the sensing requirement is sufficiently relaxed, the transmit design naturally maintains an adequate sensing accuracy to support secure communication, and thus further relaxing the sensing MSE requirement provides little additional secrecy gain.
Moreover, the performance degradation caused by the stringent sensing MSE constraint at $\Gamma_{\rm th}=0.2$ becomes less pronounced as $P_{\max}$ increases.
This is because a larger transmit-power budget provides greater flexibility to satisfy the sensing requirement while preserving the degrees of freedom available for secrecy-rate maximization.
{Taken together, the results in Figs.~\ref{fig:traj_zoom} and~\ref{fig:average} demonstrate that tightening the sensing MSE constraint improves the eavesdropper tracking accuracy.
The improved tracking accuracy, however, comes at the expense of reduced flexibility for secure communication, leading to a trade-off between sensing and secrecy performance.}
\begin{figure}[!t]
\centering
\subfloat[]{%
\includegraphics[width=0.48\linewidth]{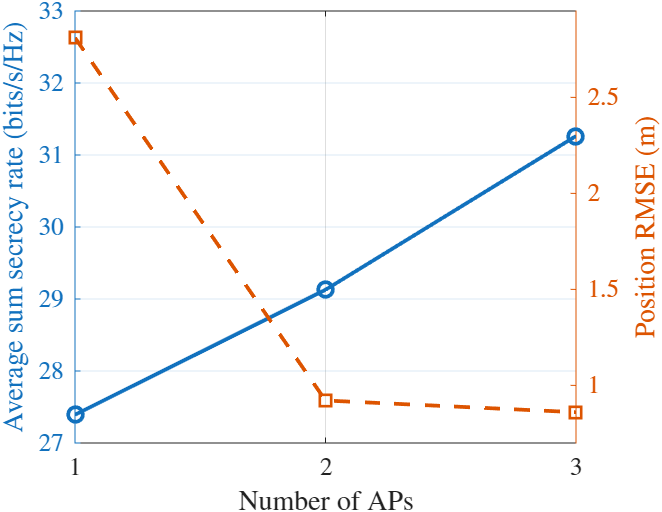}%
\label{fig:ap_sweep}%
}\hfill
\subfloat[]{%
\includegraphics[width=0.48\linewidth]{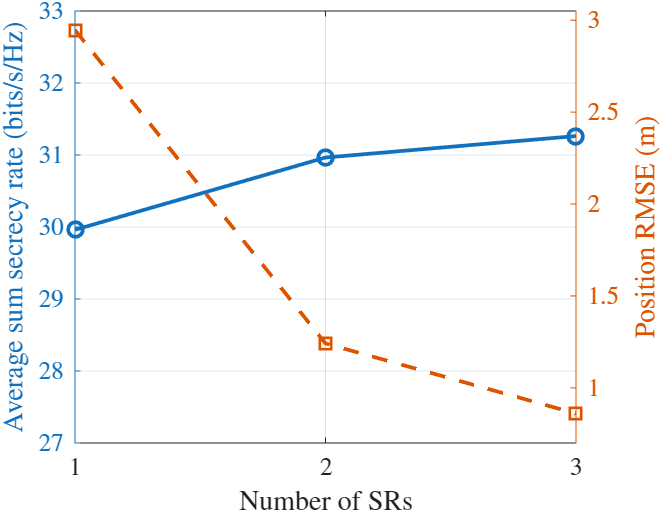}%
\label{fig:sr_sweep}%
}
\caption{Average sum secrecy rate and position RMSE versus (a) the number of APs and (b) the number of SRs under fixed total network resources.}
\label{fig:ap_sr_sweep}
\end{figure}

Fig.~\ref{fig:ap_sr_sweep} evaluates the impact of the numbers of APs and SRs on the average sum secrecy rate and position root-mean-square error (RMSE) while keeping the total network resources fixed, thereby isolating the deployment gain from the resource gain.
Specifically, when $M$ is varied from $1$ to $3$ with $V=3$, the total transmit power and the total number of transmit antennas are kept unchanged by scaling the per-AP power budget and the number of transmit antennas inversely with $M$.
Similarly, when $V$ is varied from $1$ to $3$ with $M=3$, the total number of receive antennas is fixed by evenly distributing them across the SRs.
As shown in Fig.~\ref{fig:ap_sweep}, increasing the number of APs improves the average sum secrecy rate while reducing the position RMSE.
This is because a more distributed AP deployment provides greater spatial diversity for both communication and sensing, yielding more favorable AP--user and AP--eavesdropper channel geometries as well as richer bistatic sensing links.
As shown in Fig.~\ref{fig:sr_sweep}, increasing the number of SRs also improves the average sum secrecy rate and reduces the position RMSE.
Although the SRs do not directly participate in communication transmission, deploying more SRs at different locations enhances sensing diversity and improves tracking accuracy.
The resulting reduction in the eavesdropper position uncertainty reduces the channel uncertainty, thereby enabling a higher sum secrecy rate through robust transmit design.

\begin{figure}[!t]
    \centering
    \includegraphics[width=0.6\linewidth]{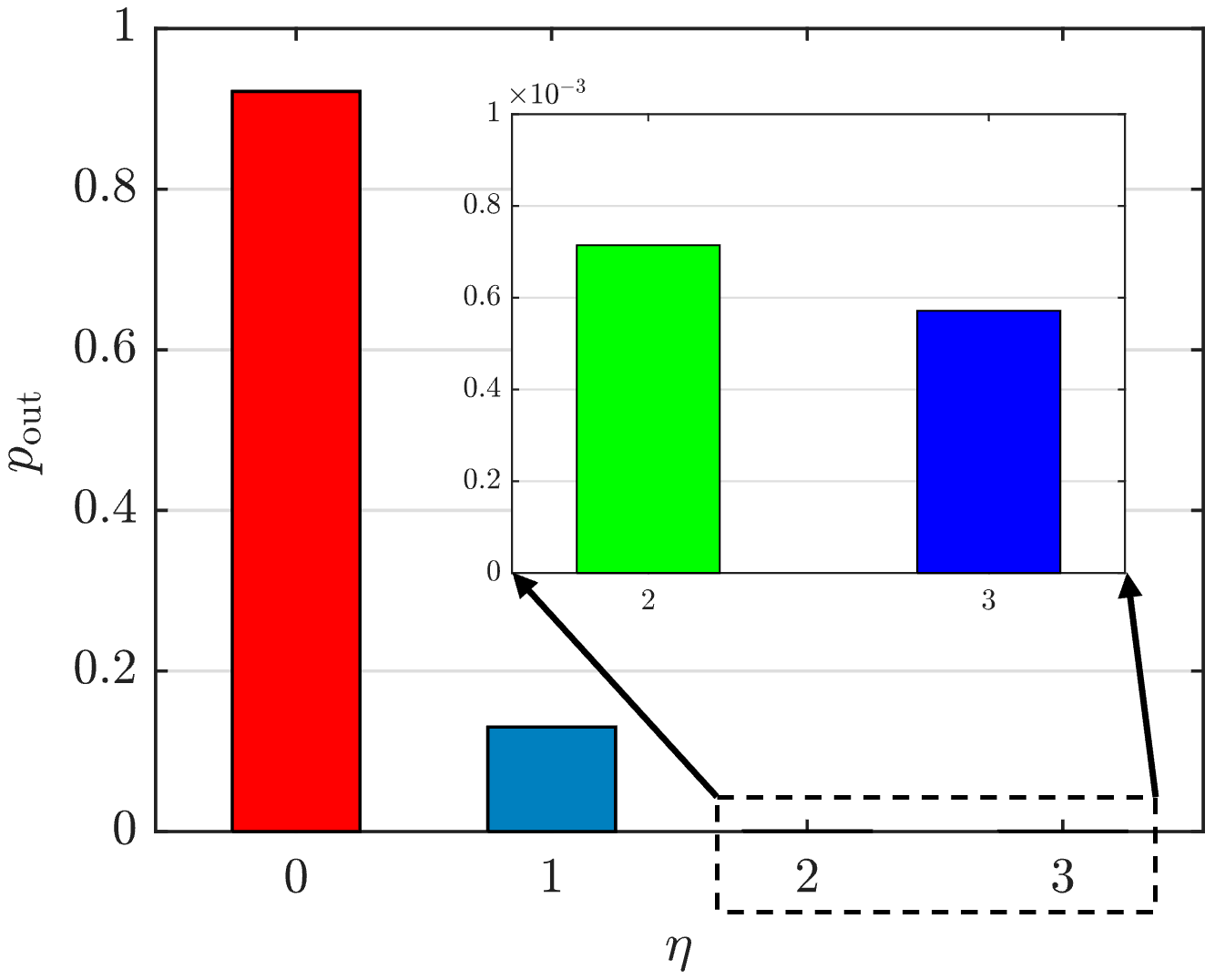}
    \caption{Outage probability versus uncertainty scaling factor $\eta$.}
    \label{fig:robustness}
\end{figure}
\begin{figure}[!t]
\centering
\subfloat[]{%
\includegraphics[width=0.48\linewidth]{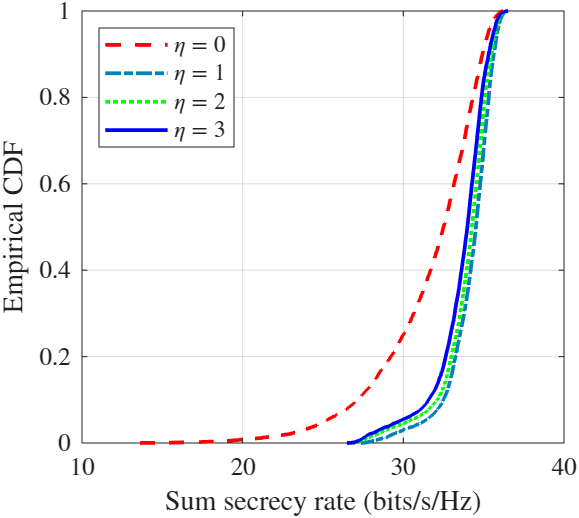}%
\label{fig:rbcomm}%
}\hfill
\subfloat[]{%
\includegraphics[width=0.48\linewidth]{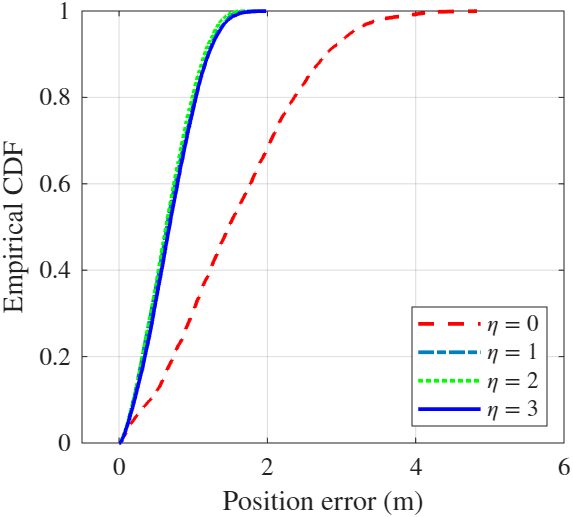}%
\label{fig:rbsen}%
}
\caption{Empirical CDFs under different uncertainty scaling factors $\eta$: (a) sum secrecy rate; (b) position error.}
\label{fig:cdfs}
\end{figure}


Fig.~\ref{fig:robustness} and Fig.~\ref{fig:cdfs} evaluate the robustness through the outage probability and the empirical cumulative distribution functions (CDFs) of the sum secrecy rate and position error, respectively, under different uncertainty scaling factors $\eta$. Fig.~\ref{fig:robustness} shows the outage probability, where an outage event is declared when the sum secrecy rate at the actual eavesdropper position is lower than the worst-case sum secrecy rate obtained from the robust optimization.
Here, $\eta=0$ corresponds to the non-robust baseline.
As shown in the figure, increasing $\eta$ substantially reduces the outage probability by accounting for a larger eavesdropper channel uncertainty region, and even $\eta=1$ achieves a significant improvement over the non-robust design.
Fig.~\ref{fig:cdfs} presents the empirical CDFs of the sum secrecy rate and position error.
As shown in Fig.~\ref{fig:rbcomm}, the non-robust design ($\eta=0$) achieves substantially lower secrecy rates than the robust designs because the resulting tracking errors lead to severe channel mismatch.
In contrast, the secrecy-rate CDFs for the robust designs differ only slightly, indicating that a moderate uncertainty region is sufficient to capture most of the robustness gain, while further enlarging the uncertainty region mainly results in a more conservative transmit design.
Similarly, Fig.~\ref{fig:rbsen} shows that the robust designs achieve noticeably smaller position errors than the non-robust baseline, whereas the position-error CDFs remain nearly unchanged for different robust settings.
Overall, these results suggest that a moderate uncertainty region is sufficient to achieve most of the robustness and tracking gains while avoiding unnecessary degradation in the achievable secrecy rate.

We next compare the proposed scheme with four structural baselines that differ in their sensing and transmission architectures.
For sensing, multistatic sensing exploits echoes generated by signals transmitted from multiple APs, whereas monostatic sensing relies only on the echo generated by each AP's own transmission.
For communication, the transmission mode is either joint transmission (JT), where all APs jointly serve the users, or coordinated beamforming (CB), where each AP serves its associated users while coordinating its beamforming design with the other APs~\cite{alexandropoulos2016advanced}.
Accordingly, we consider the following schemes:
(i) \textbf{Multi-JT}, which combines multistatic sensing with JT and corresponds to the proposed scheme;
(ii) \textbf{Mono-JT}, which combines monostatic sensing with JT;
(iii) \textbf{Multi-CB}, which combines multistatic sensing with CB;
(iv) \textbf{Mono-CB}, which combines monostatic sensing with CB; and
(v) \textbf{Single AP}, which employs a single AP for both communication and monostatic sensing while using the same total transmit power as the distributed cell-free setup. 
The single AP is equipped with $MN$ transmit antennas and $VN$ receive antennas, thereby preserving the same total antenna resources as the distributed deployment.

\begin{figure}[!t]
\centering
\subfloat[]{%
\includegraphics[width=0.48\linewidth]{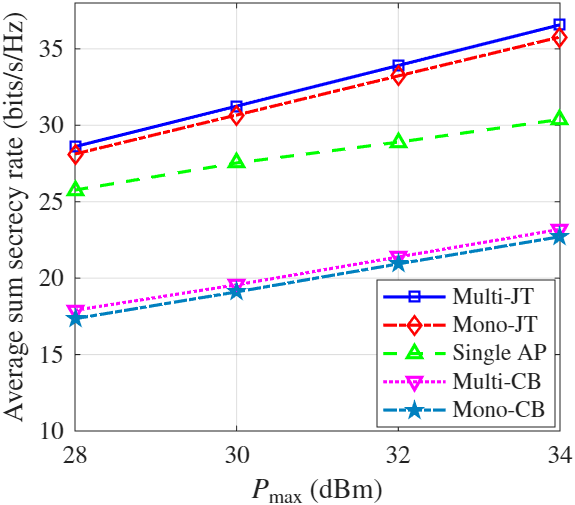}%
\label{fig:p_rate}%
}\hfill
\subfloat[]{%
\includegraphics[width=0.48\linewidth]{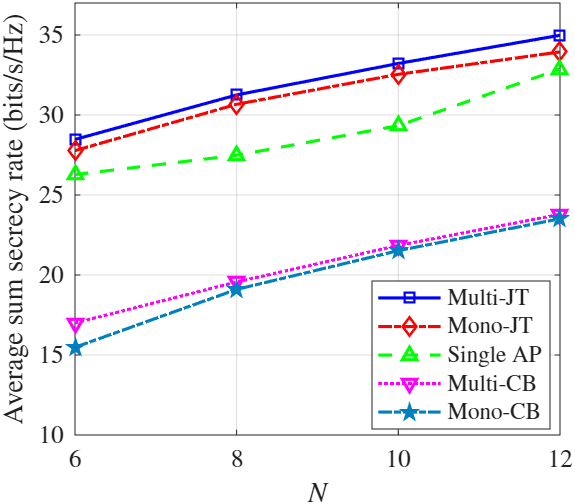}%
\label{fig:ant_rate}%
}
\caption{Average sum secrecy rate versus (a) the per-AP transmit power budget $P_{\max}$ and (b) the number of antenna elements $N$.}
\label{fig:rate_sweep}
\end{figure}

\begin{figure}[!t]
\centering
\subfloat[]{%
\includegraphics[width=0.48\linewidth]{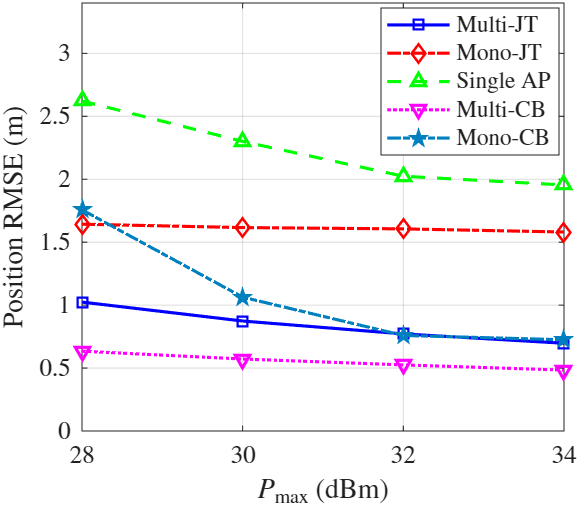}%
\label{fig:p_sen}%
}\hfill
\subfloat[]{%
\includegraphics[width=0.48\linewidth]{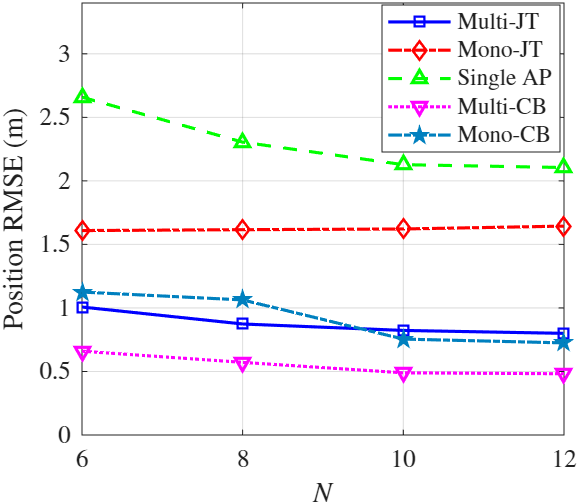}%
\label{fig:ant_sen}%
}
\caption{Position RMSE versus (a) the per-AP transmit power budget $P_{\max}$ and (b) the number of antenna elements $N$.}
\label{fig:sen_sweep}
\end{figure}

Fig.~\ref{fig:rate_sweep} evaluates the average sum secrecy rate as $P_{\max}$ and $N$ vary.
As shown in Fig.~\ref{fig:p_rate}, the average sum secrecy rate increases with $P_{\max}$.
For all considered transmit-power budgets, the JT-based schemes consistently outperform the CB-based schemes because joint transmission fully exploits cooperative beamforming across multiple APs, whereas CB leaves residual inter-cell interference by serving each user only through its associated AP.
Indeed, the CB-based schemes even underperform \textbf{Single AP}, indicating that distributed deployment alone is insufficient and that inter-AP transmission cooperation is essential for effectively utilizing distributed transmit resources.
Fig.~\ref{fig:ant_rate} exhibits a similar trend as the average sum secrecy rate increases with $N$ due to the increased array gain and spatial degrees of freedom. Moreover, for a given transmission mode, the multistatic sensing-based schemes consistently achieve higher secrecy rates than their monostatic counterparts, suggesting that improved tracking accuracy reduces the eavesdropper channel uncertainty and thereby facilitates more effective robust transmit design.
Fig.~\ref{fig:sen_sweep} evaluates the corresponding position RMSE.
As shown in Fig.~\ref{fig:p_sen} and Fig.~\ref{fig:ant_sen}, the position RMSE generally decreases as $P_{\max}$ and $N$ increase.
Moreover, the distributed schemes consistently outperform \textbf{Single AP}, and, for a given transmission mode, multistatic sensing achieves lower position RMSEs than monostatic sensing owing to the increased sensing diversity.
Interestingly, the CB-based schemes achieve comparable or even lower position RMSEs than their JT-based counterparts, indicating that the additional transmission cooperation in JT is primarily exploited to improve secrecy performance rather than tracking accuracy.
\section{Conclusion}
\label{sec:conclusion}
In this paper, we proposed a tracking-assisted robust secure transmission framework for cell-free ISAC against a mobile eavesdropper. By jointly integrating multistatic sensing, EKF-based tracking, uncertainty modeling, and robust beamforming, the proposed framework effectively incorporates mobility-induced position uncertainty into secure transmission design. Numerical results demonstrate that reliable secure transmission requires balancing tracking accuracy and secrecy performance while leveraging distributed multistatic sensing, uncertainty-aware robust beamforming, and cooperative transmission. Overall, this work highlights the potential of cell-free ISAC as a robust and scalable architecture for physical-layer security against mobile eavesdroppers.

Future work may extend the proposed framework to scenarios involving multiple mobile eavesdroppers. Another promising direction is to consider mobile legitimate users, where sensing information can be exploited to jointly account for the position uncertainties of both legitimate users and eavesdroppers in secure transmission design. It is also of interest to develop adaptive uncertainty-region design methods that dynamically adjust the uncertainty region according to the tracking accuracy and the desired level of robustness.


\appendices
\section{EKF Measurement Model}
\label{app:EKF}
As described in Section~\ref{sec:st_model}, the measurement parameters are estimated for each AP--SR pair through MF processing. 
Accordingly, the scalar measurement equations corresponding to~\eqref{eq:meas_model} are specified as
\begin{equation}
\begin{cases}
\hat{\tau}_{m,v}[l]
=
\frac{\|{\bf q}_e[l]-{\bf q}_{m}\|+\|{\bf q}_e[l]-{\bf q}_{v}\|}{c}
+n_{\tau_{m,v}[l]},
\\
\hat{\mu}_{m,v}[l]
=
\frac{f_c}{c}
\left[
\frac{\dot{\bf q}_e^T[l]({\bf q}_e[l]-{\bf q}_{m})}{\|{\bf q}_e[l]-{\bf q}_{m}\|}
+
\frac{\dot{\bf q}_e^T[l]({\bf q}_e[l]-{\bf q}_{v})}{\|{\bf q}_e[l]-{\bf q}_{v}\|}
\right]
+n_{\mu_{m,v}[l]},
\\
\sin\hat{\vartheta}_{m,v}[l]
=
\frac{x_e[l]-x_v}{\|{\bf q}_e[l]-{\bf q}_{v}\|}
+n_{\sin\vartheta_{m,v}[l]},
\end{cases}
\label{eq:meas}
\end{equation}
where $c$ and $f_c$ denote the speed of light and the carrier frequency, respectively. 
The measurement noises $n_{\tau_{m,v}[l]}$, $n_{\mu_{m,v}[l]}$, and $n_{\sin\vartheta_{m,v}[l]}$ are modeled as zero-mean Gaussian random variables with variances $\sigma^2_{\tau_{m,v}[l]}$, $\sigma^2_{\mu_{m,v}[l]}$, and $\sigma^2_{\sin\vartheta_{m,v}[l]}$, respectively.
Moreover, according to~\cite{liu2020radar}, the delay, Doppler, and angle-domain measurement noise variances can be expressed as
$\sigma^2_{\tau_{m,v}[l]}=a_\tau/\mathrm{SNR}_{m,v}[l]$,
$\sigma^2_{\mu_{m,v}[l]}=a_\mu/\mathrm{SNR}_{m,v}[l]$, and
$\sigma^2_{\vartheta_{m,v}[l]}=a_\vartheta/\mathrm{SNR}_{m,v}[l]$, respectively, where $a_\tau,a_\mu,a_\vartheta>0$ are constants determined by the system configuration, signal design, and the specific signal processing algorithm. 
Furthermore, following~\cite{liu2023securing,zhao2025sensing,wei2023integrated}, when the angle-domain noise variance $\sigma^2_{\vartheta_{m,v}[l]}$ is sufficiently small, the sine-domain measurement noise variance can be approximated using a first-order trigonometric transformation as
$\sigma^2_{\sin\vartheta_{m,v}[l]}
\approx
\cos^2\hat{\vartheta}_{m,v}[l]\sigma^2_{\vartheta_{m,v}[l]}$.
Therefore, the measurement noise covariance matrix for the AP--SR pair $(m,v)$ is given by
\begin{equation}
{\bf R}_{m,v}[l]
=
\mathrm{diag}
\left(
\sigma^2_{\tau_{m,v}[l]},
\sigma^2_{\mu_{m,v}[l]},
\sigma^2_{\sin\vartheta_{m,v}[l]}
\right).
\label{eq:meas_cov}
\end{equation}
Furthermore, for the EKF implementation, we provide the linearization of the nonlinear measurement function. 
\begin{figure*}[!t]
\begin{equation}
\frac{\partial {\bf g}_{m,v}({\bm\chi})}{\partial {\bm\chi}}=\begin{bmatrix}\frac{1}{c}\left(
\frac{x_e-x_{m}}{\|{\bf q}_e-{\bf q}_{m}\|}
+\frac{x_e-x_{v}}{\|{\bf q}_e-{\bf q}_{v}\|}\right)
&\frac{1}{c}\left(
\frac{y_e-y_{m}}{\|{\bf q}_e-{\bf q}_{m}\|}
+\frac{y_e-y_{v}}{\|{\bf q}_e-{\bf q}_{v}\|}\right)&0&0\\\frac{\partial g_{\mu_{m,v}}}{\partial x_e}&\frac{\partial g_{\mu_{m,v}}}{\partial y_e}&\frac{\partial g_{\mu_{m,v}}}{\partial \dot{x}_e}&\frac{\partial g_{\mu_{m,v}}}{\partial \dot{y}_e}\\
\frac{(y_e-y_{v})^2}{\|{\bf q}_e-{\bf q}_{v}\|^3}&\frac{-(x_e-x_{v})(y_e-y_{v})}{\|{\bf q}_e-{\bf q}_{v}\|^3}
&0&0\end{bmatrix}
\label{eq:G}
\end{equation}
\vspace{0.5em}
\hrule
\end{figure*}
Specifically, $\frac{\partial {\bf g}_{m,v}({\bm\chi})}{\partial {\bm\chi}}$ is given in~\eqref{eq:G}, where 

\begin{multline}
  \!\!\!\!\!  \frac{\partial g_{\mu_{m,v}}}{\partial x_e}=
\frac{f_c}{c}\left[
\frac{\dot{x}_e}{\|{\bf q}_e-{\bf q}_{m}\|}-\frac{\dot{\bf q}_e^T({\bf q}_e-{\bf q}_{m})(x_e-x_{m})}{\|{\bf q}_e-{\bf q}_{m}\|^3}\right.
\\+\left.
\frac{\dot{x}_e}{\|{\bf q}_e-{\bf q}_{v}\|}-\frac{\dot{\bf q}_e^T({\bf q}_e-{\bf q}_{v})(x_e-x_{v})}{\|{\bf q}_e-{\bf q}_{v}\|^3}
\right],
\label{eq:g1}
\end{multline}
\begin{multline}
    \!\!\!\!\! \frac{\partial g_{\mu_{m,v}}}{\partial y_e}=\frac{f_c}{c}\left[
\frac{\dot{y}_e}{\|{\bf q}_e-{\bf q}_{m}\|}-\frac{\dot{\bf q}_e^T({\bf q}_e-{\bf q}_{m})(y_e-y_{m})}{\|{\bf q}_e-{\bf q}_{m}\|^3}\right.
\\+\left.
\frac{\dot{y}_e}{\|{\bf q}_e-{\bf q}_{v}\|}-\frac{\dot{\bf q}_e^T({\bf q}_e-{\bf q}_{v})(y_e-y_{v})}{\|{\bf q}_e-{\bf q}_{v}\|^3}
\right],
\label{eq:g2}
\end{multline}
\begin{equation}
    \frac{\partial g_{\mu_{m,v}}}{\partial \dot{x}_e}=\frac{f_c}{c}\left[
\frac{x_e-x_{m}}{\|{\bf q}_e-{\bf q}_{m}\|}+
\frac{x_e-x_{v}}{\|{\bf q}_e-{\bf q}_{v}\|}
\right],
\label{eq:g3}
\end{equation}
\begin{equation}
\frac{\partial g_{\mu_{m,v}}}{\partial \dot{y}_e}=\frac{f_c}{c}\left[
\frac{y_e-y_{m}}{\|{\bf q}_e-{\bf q}_{m}\|}+
\frac{y_e-y_{v}}{\|{\bf q}_e-{\bf q}_{v}\|}
\right].
\label{eq:g4}
\end{equation}

\bibliographystyle{IEEEtran}
\bibliography{ref}

\end{document}